\documentclass[prd,tightenlines,showpacs,nofootinbib,eqsecnum,amsfonts,amssymb,amsmath,twocolumn]{revtex4}
\usepackage{graphicx}
\usepackage[dvips,usenames]{color}
\usepackage{dsfont}
\usepackage{bm}
\begin{document}
\newcommand{\dd}{{\rm d}}
\newcommand{\etainit}{\eta_0}
\newcommand{\ii}{{\rm i}}
\newcommand{\etal}{{\it et al.}}
\def\HH{\mathcal{H}}
\newcommand{\sig}[2]{\sigma^{#1}_{\,\,#2}}
\newcommand{\ta}[2]{\tau^{#1}_{\,\,#2}}
\def\sva{\sigma_{_{\rm V}a}}
\def\svb{\sigma_{_{\rm V}b}}
\def\svun{\sigma_{_{\rm V}1}}
\def\svdeux{\sigma_{_{\rm V}2}}
\def\stl{\sigma_{_{\rm T}\lambda}}
\def\stplus{\sigma_{_{\rm T}+}}
\def\stcross{\sigma_{_{\rm T}\times}}
\def\spar{\sigma_{_\parallel}}
\def\stplus{\sigma_{_{\rm T}+}}
\def\stcross{\sigma_{_{\rm T}\times}}
\def\Sva{\Sigma_{_{\rm V}a}}
\def\Svb{\Sigma_{_{\rm V}b}}
\def\Svun{\Sigma_{_{\rm V}1}}
\def\Svdeux{\Sigma_{_{\rm V}2}}
\def\Stl{\Sigma_{_{\rm T}\lambda}}
\def\Stplus{\Sigma_{_{\rm T}+}}
\def\Stcross{\Sigma_{_{\rm T}\times}}
\def\Spar{\Sigma_{_\parallel}}
\def\Stplus{\Sigma_{_{\rm T}+}}
\def\Stcross{\Sigma_{_{\rm T}\times}}
\newcommand{\urel}[1]{{\tt #1}}

\title{Perturbations of generic Kasner spacetimes and their stability}

\author{Lev Kofman$^{1}$}
\email{cyril.pitrou@port.ac.uk}
\author{Jean-Philippe Uzan$^{2,3,4}$}
\email{uzan@iap.fr}
\author{Cyril Pitrou$^5$}
\email{cyril.pitrou@port.ac.uk}
\affiliation{
$^1$ CITA, Univsersity of Toronto, 60 St. George Street, Toronto, Ontario, Canada, M5S 3H8\\
$^2$Institut d'Astrophysique de Paris, UMR-7095 du CNRS, 
                       Universit\'e Pierre et Marie Curie, 98 bis bd Arago, 75014 Paris, France\\
$^3$Department of Mathematics and Applied Mathematics, Cape Town University, Rondebosch 7701, South Africa\\
$^4$National Institute for Theoretical Physics (NITheP), Stellenbosch 7600, South Africa\\
$^5$Institute of Cosmology and Gravitation, Dennis Sciama Building, Burnaby Road, 
                        Portsmouth, PO1 3FX, United Kingdom.}
\pacs{98.80.-k,98.80.Cq}

\begin{abstract}
This article investigates the stability of a generic Kasner spacetime to linear
perturbations, both at late and early times. It demonstrates that the 
perturbation of the Weyl tensor diverges at late time in all cases but in
the particular one in which the Kasner spacetime is the product of a
two-dimensional Milne spacetime and a two-dimensional Euclidean space.
At early times, the perturbation of the Weyl tensor also diverges
unless one imposes a condition on the perturbations so as
to avoid the most divergent modes to be excited. 
\end{abstract}
 \date{\today}
 \maketitle
\section{Introduction}

The formalism of cosmological perturbations about a Bianchi~I universe with a scalar
field was only developed recently in Refs.~\cite{ppu1,ppu2} and in Ref.~\cite{emir0} for
the subcase of axisymmetric spacetimes. In such anisotropic inflationary models,
the shear is always dominating at early time so that the universe behaves as
a Kasner spacetime. It was realized that this anisotropic early era has an important
signature since gravity waves are amplified during this era~\cite{ppu2,emir}. Indeed, this
stage is usually short and this instability is transient.

However this has led to question the stability of a pure Kasner universe~\cite{emir}.
The Kasner spacetimes~\cite{kasner} are vacuum solutions of the Einstein field
equations. They describe universes which are spatially homogeneous and Euclidean but with an anisotropic expansion. 
They play an important role in cosmology since they are a key structure
in the discussion of the dynamics of spatially homogeneous spacetimes close 
to the singularity. Belinsky, Khalatnikov
and Lifshitz~\cite{bkl1,bkl2,bkl3} and Misner~\cite{mixmaster} investigated the nature of the cosmological
singularity by means of a Bianchi~IX model, whose temporal behaviour toward the singularity was
shown to be described by a sequence of anisotropic Kasner era. 
This has led to the mixmaster picture~\cite{mixmaster} and the idea
of the cosmic billiard~\cite{bklnew} that can offer a description
of the geometry of the universe prior to inflation.

The stability of this picture was already adressed in Ref.~\cite{bkl1}
and later  numerically in Refs.~\cite{B9stability,B9stability2,adams}.
Only recently was it revisited~\cite{emir} in light of the recent developments
concerning Bianchi~I universe perturbations~\cite{ppu1,ppu2,emir0} 
in the particular case of a Kasner universe with an axial symmetry, but
no general study of the stability to linear perturbations of a Kasner
spacetime with arbitrary exponents has been performed yet. This is
the goal of this article.

We limit our analysis to the study of the stability of the Kasner spacetime 
to linear perturbations.
Indeed, this can exhibit some instabilities but does not allow to demonstrate
the general stability when no instability is exhibited at linear order in the perturbations. For instance, 
the stability of the Minkowski spacetime to linear perturbation is an
easy exercise while the general proof of stability is extremely
challenging~\cite{christo}. Also some spacetimes may be unstable
to a particular class of perturbations while stable in other
cases. For instance, the Einstein static universe was shown to be unstable
with respect to spatially homogeneous
and isotropic perturbations~\cite{eddi} while it was then understood that
the issue was not that simple when Harrison~\cite{harri} demonstrated that
all physical inhomogeneous modes are oscillatory if the matter content
was a radiation fluid, while Gibbons~\cite{gibbons} demonstrated
the stability  against conformal metric perturbations when the matter content
was a fluid with sound speed larger that $\frac{1}{5}$. Finally~\cite{be},
it was shown that the Einstein static spacetime is
neutrally stable against small inhomogeneous vector and tensor perturbations and neutrally
stable against adiabatic scalar density inhomogeneities as long as the
condition exhibited by Gibbons holds, and unstable otherwise. Our analysis is
somehow simplified by the fact that the Kasner spacetimes are empty so that
we need no assumption on the matter content or on the type of perturbations
considered. Once the evolution of the linear perturbation is determined, in a gauge
invariant formalism to ensure that no gauge mode can spoil the conclusion,
we compute the invariants of the metric and compare them to those
of the background. Again, since the Kasner spacetime is empty, only
the square of the Weyl tensor has to be considered. This ensures that
the conclusion is not affected by the choice of gauge or coordinates system.

In this article, we consider the perturbation theory around generic Kasner
spacetimes, as described in \S~\ref{sec1},
in order to discuss their stability. First, in \S~\ref{sec2}, starting from
the general formalism we developed in Refs.~\cite{ppu1,ppu2}, we show that for a vacuum
spacetime only two perturbations are propagating, as expected. We then
exhibit some asymptotic analytical solutions for the behaviour of
the perturbations in \S~\ref{sec3} which allow to discuss the
stability at late and early times in the generic case.
In \S~\ref{secaxial}, we consider the axially symmetric case considered
in Ref.~\cite{emir0}, mostly in order to discuss the agreement with this
previous analysis.

\section{Background spacetime}\label{sec1}

\subsection{Definitions}\label{sec2-a}

We consider a Kasner spacetime, that is a Bianchi~I universe
with metric usually written in terms
of the cosmic time $t$ as
\begin{equation}\label{metricB1}
  \dd s^2 = -\dd t^2 + S^2(t)\gamma_{ij}(t)\dd x^i\dd x^j
\end{equation}
for the vacuum, where $S$ is the volume averaged scale factor
and $\gamma_{ij}(t)$ the metric on the constant time hypersurfaces.
The vacuum Einstein equations 
imply that $\dot H=-3H^2$, 
from which we deduce that the averaged Hubble parameter is
\begin{equation}
 H \equiv \frac{\dot S}{S} = \frac{1}{3t}\,.
\end{equation}
It is easily integrated to deduce that $S(t) =S_0 (t/t_0)^{\frac13}$ and we 
can always set $S_0=1$.

\begin{figure}
 \includegraphics[width=8.5cm]{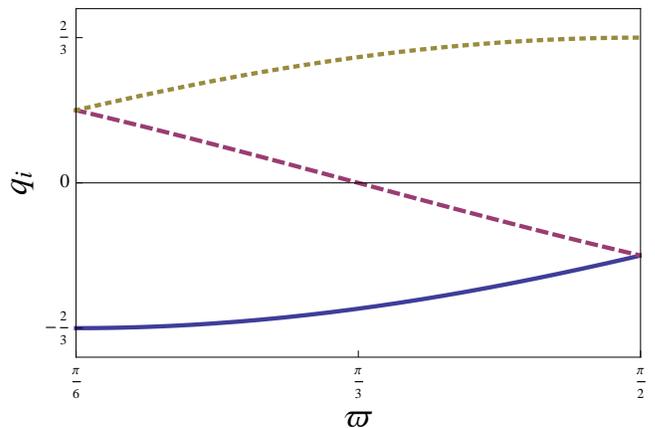}
 \caption{Evolution of $q_i$ with $\varpi$
  ($q_1$ in solid line, $q_2$ in dashed line and $q_3$
  in dotted line). We can choose $\varpi\in[\frac{\pi}{6},\frac{\pi}{2}]$ so that
 $p_1\leq p_2\leq p_3$.}\label{f1}
\end{figure}

The Kasner metric can then be shown to be of the form
\begin{equation}
 \dd s^2 = -\dd t^2 + \sum_i \left[\left(\frac{t}{t_0}\right)^{p_i}\dd x^i\right]^2,
\end{equation}
with the coefficients $p_i$ satisfying the constraints
\begin{equation}\label{Eqconstraintspi}
 \sum_i p_i =\sum_i p_i^2 = 1,
\end{equation}
that are fulfilled if
\begin{equation}\label{paravarpi}
 p_i=\frac{2}{3}\sin\varpi_i + \frac{1}{3},\qquad
 \varpi_i\equiv \varpi - \frac{2\pi}{3}i.
\end{equation}
$\varpi$ varies in an interval of length $2\pi/3$
and we always have $p_1 \le (p_2,p_3)$ if $\varpi\in [-\pi/6,\pi/2]$
but it is enough to let it vary in $\varpi\in ]\pi/6,\pi/2[$, which always ensures
that $p_1\leq p_2\leq p_3$ (see Fig.~\ref{f1}). The two limiting cases
$\varpi=\frac{\pi}{6}$ or $\varpi=\frac{\pi}{2}$
for which $p_i=\lbrace{-\frac13,\frac23,\frac23\rbrace}$
and $p_i=\lbrace{0,0,1\rbrace}$ respectively,
correspond to space-times with an extra axial symmetry so that
$p_2=p_3$ or $p_2=p_1$.

It is thus clear that the spatial metric is explicitely given by
\begin{equation}
 \gamma_{ij}(t)= 
 {\rm diag}\left[\left(\frac{t}{t_0}\right)^{2q_i}\right]
 \quad\hbox{with}
\quad
q_i=p_i-\frac{1}{3}
\end{equation}
so that
\begin{equation}
 \sum_i q_i=0,\qquad  \sum_i q_i^2=\frac{2}{3}\,.
\end{equation}
The $q_i$ are explicitely given by
\begin{equation}
q_i=\frac{2}{3}\sin\varpi_i.
\end{equation}
We also define the useful parameterisation
\begin{equation}
Q_i \equiv \frac{3}{2} q_i = \sin \varpi_i
\end{equation}
such that $Q_i \in [-1,1]$.
We define the conformal time $\eta$ from $\dd t \equiv S \dd\eta$, and
the metric~(\ref{metricB1}) can then be recast as
\begin{equation}\label{metricconforme}
 \dd s^2 = S^2(\eta)\left[-\dd \eta^2+ \gamma_{ij}(\eta)\dd x^i\dd x^j\right].
\end{equation}
From the previous analysis, we have $\eta =\frac{3}{2}t_0
(t/t_0)^{\frac23}$ so that we set
\begin{equation}
 S(\eta) = \sqrt{\frac{\eta}{\eta_0}}\quad\hbox{with}\quad
 \eta_0=\frac{3}{2}t_0
\end{equation}
and also 
\begin{equation}\label{EqRelteta}
t = \frac{2}{3} S \eta.
\end{equation}
We easily deduce that
\begin{equation}
 \gamma_{ij} ={\rm diag}\left[\left(\frac{\eta}{\eta_0}\right)^{3q_i}\right], \quad
  \gamma^{ij} ={\rm diag}\left[\left(\frac{\eta}{\eta_0}\right)^{-3q_i}\right]. 
\end{equation}
We define the comoving Hubble parameter by $\mathcal{H}\equiv S'/S$, where a prime refers to a
derivative with respect to the conformal time so that
\begin{equation}
 \HH =S H= \frac{1}{2\eta}\,.
\end{equation}
The shear tensor is defined as $\sigma_{ij} \equiv
\frac{1}{2}\gamma_{ij}'$ and is traceless ($\gamma^{ij}
\sigma_{ij}\equiv \sigma^i_i=0$) so that
\begin{equation}\label{defsigmaij}
 \Sigma_{ij} \equiv \frac{\sigma_{ij}}{\HH}
   =3{\rm diag}\left[q_i\left(\frac{\eta}{\eta_0}\right)^{3q_i}\right]
\end{equation}
and thus $\Sigma^i _j=3{\rm diag}[q_i]$ so that
\begin{equation}\label{propkasner}
 \Sigma^2 = 6, \qquad
 2S^2\HH=\eta_0^{-1}
\end{equation}
and the Einstein equations imply that the shear satisfies
\begin{equation}\label{ein-tt}
 \left(\sigma^i_j \right)' + 2\HH \sigma^i_j=0.
\end{equation}
We also define the shear in cosmic time by 
\begin{equation}
\hat \sigma_{ij} \equiv  \frac{1}{2}\frac{\dd\gamma_{ij}}{\dd t} = \frac{\sigma_{ij}}{S}\,.
\end{equation}

\subsection{Weyl Tensor}

The Kasner spacetime is a vacuum solution of Einstein equations and
this implies that
\begin{equation}
 R_{\mu\nu}=0, \quad R =0, \quad R_{\mu\nu\rho\sigma} = C_{\mu\nu\rho\sigma}.
\end{equation}
It is thus only characterized by its Weyl tensor, both at the background and
perturbation level. The Weyl tensor has just two types of
non-vanishing components which are the $C^{0i}_{\phantom{0i}0j}$ and
the $C^{0i}_{\phantom{0i}jk}$. The former are related to the
components of the type $C^{ij}_{\phantom{ij}kl}$ (thanks to the
traceless conditions) and to the electric part of the Weyl tensor. The
latter are related to its magnetic part, and it vanishes for the
background Kasner spacetime. The only non-vanishing components of the
background Weyl tensor are thus expressed as
\begin{eqnarray}
 C^{0i}_{\phantom{0i}0j} &=& -H \hat \sigma^i_{j} -\frac{1}{3}\hat
 \sigma^2\delta^i_{j} + \hat \sigma^i_{k}\hat \sigma^k_j,\\
 C^{ij}_{\phantom{ij}kl}&=&
 \frac{1}{3}\hat \sigma^2\gamma^i_{[k}\gamma_{l]}^{j} +
 2\hat \sigma^i_{[k}\hat \sigma_{l]}^j\nonumber\\
&&+ 2H\left(\gamma^i_{[k}\hat \sigma_{l]}^j + \hat \sigma^i_{[k}\gamma_{l]}^j   \right),
\end{eqnarray}
where $[ij]$ refers to the antisymmetrisation of indices such that for an antisymmetric
tensor $X_{ij}=X_{[ij]}$ (and similarly later on, $(ij)$ refers to the
symmetrisation of indices such that for a symmetric tensor $X_{ij}=X_{(ij)}$).
These Weyl components are explicitely given by
\begin{eqnarray}
 C^{0i}_{\phantom{0i}0j} &=&-\frac{1}{3t^2}\left(q_i+\frac{2}{3} -3q_i^2\right)\delta^i_{j},\\
               &=&-\frac{1}{t^2}p_i\left(1-p_i\right)\delta^i_{j},\\
 C^{ij}_{\phantom{ij}kl}&=& \frac{2}{t^2}p_ip_j \delta^i_{[k}\delta_{l]}^j,
\end{eqnarray}
and it is easy to notice explicitely that these two types of components are not independent and are related thanks to Eq.~(\ref{Magicrelation123}). We thus obtain that $C^2 \equiv C_{\mu\nu\rho\sigma}C^{\mu\nu\rho\sigma}$ is given by
\begin{eqnarray}
 C^2 & = & 
 12{\cal C}^2\sin^2\left(\frac{3\varpi}{2}+\frac{\pi}{4}\right)\nonumber\\
         &=& -3^4{\cal C}^2 p_1 p_2 p_3\nonumber\\
         &=&  3^4{\cal C}^2 p_I^2(1-p_I)\nonumber\\
         &=& 3{\cal C}^2 \left(1+3 q_I\right)^2\left(2-3 q_I\right)\nonumber\\
         &=& 6{\cal C}^2 \left(1+2 Q_I\right)^2\left(1-Q_I\right)\label{valC}
\end{eqnarray} 
where in the last equalities, $I$ can take any of the values
$1,2,3$.  The function ${\cal C}^2$ is given by [using Eq.~(\ref{EqRelteta})]
\begin{equation}
{\cal C}^2 = \left(\frac{2}{3 t}\right)^4 = \frac{1}{(S\eta)^4}\,.
\end{equation}

The Weyl tensor and his algebraic properties are used to classify spacetimes
by their Petrov type. Defining the tracefree symmetric rank-2 tensor $Q_{ij}= -E_{ij} - {\rm i}H_{ij}$
in terms of the electric ($E_{ij}$) and magnetic ($H_{ij}$) parts of the Weyl tensor, provides a way to classify
the Weyl tensors~\cite{petrov}. For a Kasner spacetime, since the
magnetic part of the Weyl tensor vanishes we obtain
\begin{eqnarray}
Q^i_j = -E^i_j = C^{0i}_{\phantom{0i}0j}=-\frac{1}{t^2}p_i\left(1-p_i\right)\delta^i_j.
\end{eqnarray}
The eigenvalues $\lambda_i$ of $Q^i_j$ define its Petrov type. They are all real
and different so that a generic Kasner spacetime ($\varpi\not=\frac{\pi}{6},\frac{\pi}{2}$) has Petrov type I
while the two axially symmetric solutions with degenerate eigenvalues
are of type O when $\varpi=\frac{\pi}{2}$ and of type D when $\varpi=\frac{\pi}{6}$~\cite{petrov}. 

\section{Perturbation theory in a Kasner spacetime}\label{sec2}

The Kasner spacetime being a particular case of Bianchi universe,
the study of the evolution of the perturbations derives simply
from our previous formalism~\cite{ppu1,ppu2} that we
specialize to the vacuum, using that $S''/S = -1/(4\eta^2)$
so that $\HH'=-2\HH^2 $ and the property~(\ref{propkasner}).

\subsection{Generalities}

Following the formalism developed in Refs.~\cite{ppu1,ppu2}, we consider the
general metric of an almost Bianchi~I spacetime,
\begin{eqnarray}\label{dmet1}
  \dd s^{2}&=&S^{2}\left[-\left(1+2A\right)\dd\eta^{2}+2\bar{B}_{i}\dd x^{i}\dd\eta\right.\nonumber\\
    &&\qquad\qquad\left. +\left(\gamma_{ij}+h_{ij}\right)\dd x^{i}\dd x^{j}\right].
\end{eqnarray}
$\bar{B}_{i}$ and $h_{ij}$ can be further decomposed into 
scalar, vector and tensor components as
\begin{eqnarray}\label{tens-decomp}
\bar{B}_{i} & = & \partial_{i}B+B_{i}\,, \\
h_{ij} & \equiv & 2C\left(\gamma_{ij}+\Sigma_{ij}\right)+2\partial_{i}\partial_{j}E+2\partial_{(i}E_{j)}+2 \tilde E_{ij}\,,\nonumber
\end{eqnarray}
with
\begin{equation}
 \partial_iB^{i}=0=\partial_i E^i, \quad
 \tilde E_i^i=0=\partial_i\tilde E^{ij}.
\end{equation}
We then construct the gauge invariant quantities and define the 
conformal Newtonian gauge by the conditions
\begin{equation}
 B=B^i=E=0\,,
\end{equation}
so that
\begin{equation}
  A=\Phi\,,\qquad C=-\Psi\,,\qquad \Phi^i=-(E^i)'\,.
\end{equation}
We also introduce the extremely useful variable~\cite{pubook}
\begin{equation}\label{e:defX}
 X \equiv \Phi+\Psi+\left(\frac{\Psi}{\HH}\right)'.
\end{equation}
The tensor variable $\tilde E_{ij}$ is readily gauge invariant.
At this stage, we are left with $10-4=6$ degrees of freedom:
two scalars ($\Phi$ and $\Psi$), two vectors ($\Phi^i$) and two
tensors ($ \tilde E_{ij}$). Contrary to the perturbation
theory around a spatially homogeneous and
isotropic spacetime~\cite{pubook}, the three types of perturbations
do not decouple.

As we shall now see, the perturbation equations will exhibit
four constraints so that only two dynamical degrees of freedom related
to the gravity waves remain, as in Minkowski or de Sitter spacetimes.

\subsection{Mode decomposition}\label{SecModeDecomposiiton}

The perturbation equations are conveniently written in Fourier
space and we decompose any quantity in Fourier modes as follows. 
Using the Cartesian comoving coordinates system $\{x^i\}$ on the constant time hypersurfaces, we decompose any scalar function as
\begin{equation}\label{fft}
 f\left(x^j,\eta\right)= \int\frac{\dd^3k_i}{\left(2\pi\right)^{\frac32}}
         \, f\left(k_i,\eta\right)\,\hbox{e}^{\mathrm{i}k_ix^i}\,.
\end{equation}
The comoving wave co-vectors $k_i$ are constant, $k_i'=0$. We now define
$k^i\equiv\gamma^{ij}k_j$ which is now time-dependent and explicitely given by
\begin{equation}\label{deff2}
 k^i = k_i\left(\frac{\eta}{\eta_0}\right)^{-2 Q_i}
\end{equation}
so that 
\begin{equation}\label{deff2b}
 k^2\equiv k_i k^i = \sum_i x_i^2,\qquad
x_i\equiv k_i\left(\frac{\eta}{\eta_0}\right)^{-Q_i}\,.
\end{equation}
For any mode $k_i$, a basis $\lbrace e_1,e_2\rbrace$ of the subspace perpendicular
to $k_i$ can be constructed from the natural Cartesian basis as
(see Appendix A of Ref.~\cite{ppu2} for the details of the construction)
\begin{equation}
 e^i_1 = \left(\frac{\eta}{\eta_0}\right)^{-Q_i}\omega_1^i,\qquad
 e^i_2 = \left(\frac{\eta}{\eta_0}\right)^{-Q_i}\omega_2^i
\end{equation}
with
\begin{equation}\label{euler4}
 \omega_1^i=\left(\begin{array}{c}\cos\gamma\cos\beta\cos\alpha-\sin\alpha\sin\gamma\\
                             \cos\gamma\sin\alpha+\cos\alpha\cos\beta\sin\gamma\\
                             -\cos\alpha\sin\beta \end{array}\right),
\end{equation}
\begin{equation}\label{euler5}
  \omega_2^i =\left(\begin{array}{c}-\cos\gamma\cos\beta\sin\alpha-\cos\alpha\sin\gamma\\
                             \cos\gamma\cos\alpha-\sin\alpha\cos\beta\sin\gamma\\
                             \sin\alpha\sin\beta \end{array}\right).
\end{equation}
The three Euler angles $(\alpha,\beta,\gamma)$ depend on time and are explicitely given by
\begin{equation}\label{euler1}
 \sin\gamma\sin\beta=\frac{x_2}{k},\quad
  \cos\gamma\sin\beta=\frac{x_1}{k}
\end{equation}
so that $\tan\gamma=x_2/x_1=(k_2/k_1)(\eta/\eta_0)^{(Q_1-Q_2)}$
and
\begin{equation}\label{euler2}
\cos\beta = \frac{x_3}{k}\,.
\end{equation}
The previous relations determine $\beta(\eta)$ and $\gamma(\eta)$.
The requirement that $\lbrace e_1,e_2\rbrace$ remains
orthogonal to $k_i$ imposes that
\begin{equation}\label{euler3}
\alpha' = -\cos\beta \gamma',
\end{equation}
which can be rewritten as
\begin{equation}\label{evo-alpha2}
\alpha' = \frac{(Q_2-Q_1)}{\eta}\frac{x_3}{k}\frac{1}{\frac{x_1}{x_2}+\frac{x_2}{x_1}}\,,
\end{equation}
which determines $\alpha(\eta)$ up to an integration constant.

The vector and tensor modes can then be decomposed respectively
as
\begin{equation}\label{dec_vector}
 \Phi_i(k_i,\eta)=\sum_{a=1,2}\Phi_a(k_i,\eta) \, e_i^a(\hat k_i)\, ,
\end{equation}
and
\begin{equation}\label{dec_tensor}
 \tilde E_{ij}(k_i,\eta)=\sum_{\lambda=+,\times} E_\lambda(k_i,\eta)
 \, \varepsilon_{ij}^\lambda(\hat k_i)
\end{equation}
where the polarisation tensors have been defined as
\begin{equation}\label{defespilonij}
 \varepsilon_{ij}^\lambda \equiv \frac{e_i^1e_j^1 - e_i^2e_j^2}{\sqrt{2}}\delta^\lambda_+
 + \frac{e_i^1e_j^2 + e_i^2e_j^1}{\sqrt{2}}\delta^\lambda_\times.
\end{equation}

\subsection{Shear components}

The perturbation equations in Fourier space involve the components
of the decomposition of the shear on the basis $\lbrace \hat k,e_1,e_2\rbrace$.
Since it is a symmetric trace-free tensor, it can be decomposed as 
\begin{eqnarray}\label{decshear}
 \sigma_{ij} &=& \frac{3}{2}\left(\hat k_i\hat k_j-\frac{1}{3}\gamma_{ij}\right)\spar
 + 2\sum_{a=1,2}\sva \,\hat k_{(i}e^a_{j)}\nonumber\\
&& + \sum_{\lambda=+,\times}\stl\,\varepsilon^\lambda_{ij}\,.
\end{eqnarray}
This decomposition involves 5 independent components of the shear in a basis adapted to the
wavenumber $k_i$.  We must stress however that $(\spar,\sva,\stl)$ must not be interpreted as the Fourier
components of the shear, even if they explicitely depend on $k_i$. This dependence arises from the
local anisotropy of space. A similar decompositon for $\Sigma_{ij}$
defines the coefficients $(\Spar,\Sva,\Stl)$ and we obtain from
Einstein equation in vacuum that it must satisfy the constraint
\begin{equation}\label{sumsurs}
 6=\Sigma^2 \equiv \frac{\sigma_{ij}\sigma^{ij}}{\HH^2}
          = \frac{3}{2}\Spar^2 + 2\sum_a\Sva^2 + \sum_\lambda\Stl^2\,,
\end{equation}
which is independent of $k_i$. 

The equation of evolution of the shear~(\ref{ein-tt}) implies that
\begin{eqnarray}
 &&\Spar'=-2\HH\sum_a\Sigma_{_{\rm V}a}^2 \label{background_sigma}\\
 &&\Sva' = \HH\left(\frac{3}{2}\Sva\Spar
  -\sum_{b,\lambda}\Svb\Stl\mathcal{M}_{ab}^\lambda\right), \label{background_sigma2}\\
 &&\Stl'= 2\HH\sum_{a,b}\mathcal{M}_{ab}^\lambda\Sva\Svb \label{background_sigma3}\,,
\end{eqnarray}
where the matrix $\mathcal{M}_{ab}^\lambda$ is defined by
\begin{equation}
 {\cal M}_{ab}^\lambda = \frac{1}{\sqrt{2}}
      \left(\begin{array}{cc}
             1 &0\\0&-1
             \end{array}\right)\delta^\lambda_+
             +
     \frac{1}{\sqrt{2}}
      \left(\begin{array}{cc}
             0 &1\\1&0
             \end{array}\right)\delta^\lambda_\times\,.
\end{equation}
The equation~(\ref{decshear}) can be inverted to get
\begin{eqnarray}
\Spar &=& 2\sum_i Q_i \frac{x_i^2}{k^2},\label{s1}\\
\Sva&=&2\sum_i Q_i\frac{x_i}{k}\omega^i_a,\label{s2}\\
\Stl&=&2\sum_i Q_i\left(\frac{\eta}{\eta_0}\right)^{2 Q_i}\varepsilon_\lambda^{ii}\label{s3}.
\end{eqnarray}
The last equation implies, using Eq.~(\ref{defespilonij}), that
\begin{eqnarray}
 \Stplus &=& \sqrt{2}\sum_i Q_i\left[(\omega_1^i)^2-(\omega_2^i)^2 \right]\label{s4}\\
 \Stcross &=& 2\sqrt{2}\sum_i Q_i\omega_1^i\omega_2^i.\label{s5}
\end{eqnarray}

\subsection{Perturbation equations}

We use the equations derived in Refs.~\cite{ppu1,ppu2} when applied
to the particular case of a vacuum solution (so that $v\rightarrow0$).

The two scalar perturbations are explicitely obtained from the tensor modes
as
\begin{equation}\label{e:temp2}
 X = \frac{1}{2-\Spar} \sum_{\lambda} \Stl E_{\lambda}\, .
\end{equation}
We then deduce that $\Psi$ is given by
\begin{eqnarray}\label{magic_equation}
 k^2\Psi &=& -\HH X'  \\
 &&-  \frac{4\HH^2}{2-\Spar} 
 \left(\sum_{a} \Sva^2X - \sum_{a,b,\lambda}{\cal M}^{\lambda}_{ab}
   \Sva \Svb E_{\lambda}\right),\nonumber
\end{eqnarray}
while the second Bardeen potential is then given by
\begin{eqnarray}\label{e:scal2b}
  k^2\Phi  = k^2\left(4- \Spar\right)\Psi + 3\HH X'.
\end{eqnarray}
The vector mode is then given by
\begin{equation}\label{eqphia}
 \Phi_a  =
   -2\mathrm{i} \frac{\HH}{k}\sum_\lambda\left[
   2\sum_b {\cal M}_{ab}^\lambda \Sigma_{_{\rm V}b}
    -\Stl\frac{{\Sigma}_{_{\rm V}a}}{2-\Spar}
   \right] E_\lambda.
\end{equation}

This shows that the scalar and vector modes are obtained algebraically
from the tensor modes, which are the only degrees of freedom that propagate.
Using the shorthand notation $(1-\lambda)$ for the opposite
polarisation of $\lambda$, i.e. meaning that if $\lambda=+$, then $(1-\lambda)=\times$, and
vice-versa, and introducing 
\begin{equation}\label{e:defms}
 \mu_\lambda \equiv S E_\lambda,
\end{equation}
we have
\begin{widetext}
\begin{eqnarray}\label{eqformu}
\mu_{\lambda}''+\left(k^{2}+\frac{1}{4\eta^2}\right)\mu_{\lambda} & = & 
 2\HH^2\left(\Sigma_{_{\rm T} (1-\lambda)}^{2}\mu_{\lambda}
-\Sigma_{_{\rm T}+}\Sigma_{_{\rm T}\times}\mu_{\left(1-\lambda\right)}\right)
 +2\HH \sum_{\nu}\left(\frac{\Sigma_{_{\rm T}\nu} \Sigma_{_{\rm
T}\lambda}}{2-\Spar}\right)'\mu_{\nu}+\HH \Spar'\mu_{\lambda}.
\end{eqnarray} 
\end{widetext}
The general solution of this linear equation can always be obtained in terms of
four transfer functions as
\begin{equation}\label{e338}
E_\lambda(k_i,\eta)=T^{\uparrow}_{\lambda \lambda'}(k_i,\eta) a_{\lambda'}(k_i,\etainit)+T^{\downarrow}_{\lambda \lambda'}(k_i,\eta) b_{\lambda'}(k_i,\etainit),
\end{equation}
where $ a_{\lambda'}(k_i,\etainit)$ and $b_{\lambda'}(k_i,\etainit)$ are the initial conditions.
The notations $\downarrow$ and $\uparrow$ refer respectively to a decaying and a growing mode.
When the two polarisations are decoupled (which is the case asymptotically at early and late times, 
as we whall see below) then we only have two transfer functions to consider
$T^{\uparrow}_{\lambda \lambda'}=T^{\uparrow}_{\lambda}
\delta_{\lambda \lambda'}$ and $T^{\uparrow}_{\lambda \lambda'}=T^{\downarrow}_{\lambda}
\delta_{\lambda \lambda'}$
since the transfer of power from one polarisation to the other is negligible.
Similarly we define transfer functions for $\Psi$, $X$ and $\Phi_a$ by
\begin{equation}
\Psi(k_i,\eta)=T^{\uparrow}_{\Psi,\lambda}(k_i,\eta)
a_{\lambda}(k_i,\etainit)+T^{\downarrow}_{\Psi,\lambda}(k_i,\eta)
b_{\lambda}(k_i,\etainit),
\end{equation}
\begin{equation}
X(k_i,\eta)=T^{\uparrow}_{X,\lambda}(k_i,\eta)
a_{\lambda}(k_i,\etainit)+T^{\downarrow}_{X,\lambda}(k_i,\eta)
b_{\lambda}(k_i,\etainit),
\end{equation}
\begin{equation}
\Phi_a(k_i,\eta)=T^{\uparrow}_{\Phi_a,\lambda}(k_i,\eta) a_{\lambda}(k_i,\etainit)+T^{\downarrow}_{\Phi_a,\lambda}(k_i,\eta) b_{\lambda}(k_i,\etainit).
\end{equation}

\subsection{Summary}

In conclusion, for a given model, one can determine
$(\Spar,\Sva,\Stl)$ for each mode $k_i$ and then solve
Eq.~(\ref{eqformu}) for the gravitational waves. One
can then deduce $\Phi$, $\Psi$ and $\Phi_a$
algebraically. As expected, only two degrees
of freedom can propagate. The existence of the vector and tensor modes arises from
the fact that isotropy is violated so that the SVT modes do not decouple.
The Kasner case, being a vacuum
solution, is thus simpler than the Bianchi~I
case studied in Refs.~\cite{ppu1,ppu2}.

\section{Stability analysis}\label{sec3}

\subsection{Generalities}

In full generality, the stability analysis requires first to solve Eq.~(\ref{eqformu})
to determine the four transfer functions defined in Eq.~(\ref{e338}). This allows to
evaluate the effect of the perturbations on the square of the Weyl tensor, which is by
construction independent of the choices of gauge and coordinates system.
We thus introduce
\begin{equation}
 \Xi\equiv \frac{C^2}{\bar C^2}
\end{equation}
where $\bar C^2$ is the background value of $C^2$ given in Eq.~(\ref{valC}) 
such that at the background level $\bar \Xi=1$, and at the perturbed
level up to second order we have when $\bar C^2 \neq 0$
\begin{eqnarray}
\delta^{(1)} C^2 &=& \bar C^2 \Xi^{(1)}\,,\\
\delta^{(2)} C^2 &=& \bar C^2 \Xi^{(2)}\,.
\end{eqnarray}
The perturbation of $\Xi$ at first order in the
perturbations, $\Xi^{(1)}({\bf x},\eta)$, can be decomposed in Fourier modes as
\begin{equation}
 \Xi^{(1)}({\bf x},\eta)=\int \frac{\dd^3 {\bf k}}{(2 \pi)^{\frac32}}\Xi^{(1)}({\bf k},\eta)
 \hbox{e}^{\ii {\bf k}.{\bf x}}.
\end{equation}
In order to assess the behaviour of $C^2$ at early and late time, we assume that
at an arbitrary initial time $\etainit$ the initial conditions for the two modes
defined in Eq.~(\ref{e338}) are such that (1) they are not correlated,
\begin{equation}
\langle a_\lambda({\bf k}) b_{\lambda'}^\star({\bf k}') \rangle = 0,
\end{equation}
and that (2) they have the same initial power spectrum,
\begin{equation}
\langle a_\lambda({\bf k}) a_{\lambda'}^\star({\bf k}') \rangle =
\delta^3({\bf k}-{\bf k}')\delta_{\lambda \lambda'}P_{\rm init}(k_i),
\end{equation}
\begin{equation}
\langle b_\lambda({\bf k}) b_{\lambda'}^\star({\bf k}') \rangle =
\delta^3({\bf k}-{\bf k}')\delta_{\lambda \lambda'} P_{\rm init}(k_i).
\end{equation}
The initial power spectrum is an unknown function of the comoving 
wave-vector $k_i$. It is clear that 
\begin{equation}
 \langle  \Xi^{(1)}({\bf x},\eta)  \rangle = 0
\end{equation}
at all time. This is indeed not the case for the perturbation of $\Xi$
at second order in perturbations, $\Xi^{(2)}({\bf x},\eta) $, since it is quadratic.
The previous definitions allow to compute that at lowest order
$\langle \Xi({\bf x},\eta) \rangle =\langle \Xi^{(2)}({\bf x},\eta) \rangle$ with
\begin{equation}\label{e47}
\langle \Xi^{(2)}({\bf x},\eta) \rangle = \int  \frac{\dd^3 {\bf k} }{(2 \pi)^3} P_{\rm
  init}(k_i) \sum_\lambda \Xi^{(2)}_\lambda(\bf k,\eta)
\end{equation}
and we define
\begin{equation}
\Xi^{(2)}({\bf k},\eta) \equiv \sum_\lambda \Xi^{(2)}_\lambda({\bf k},\eta).
\end{equation}
The function $\Xi^{(2)}_\lambda(\bf k,\eta)$ can in principle be
expressed in terms of the transfer functions that appear in Eq.~(\ref{e338})
and in terms of the coefficients $\Sigma_\parallel({\bf k},\eta)$...

From a numerical point of view, the situation is thus clear but very time consuming.
It is also unnecessary to solve the evolution equations in their full generality
since we are interested in the asymptotic behaviour of $\langle \Xi^{(2)}({\bf x},\eta) \rangle$
when $\eta/\etainit\rightarrow+\infty$ or $\eta/\etainit\rightarrow 0$. In these two regimes,
the wave numbers tend to focus along a principal axis since [see Eq.~(\ref{deff2b})]
\begin{equation}\label{klarge}
 k\rightarrow k_1\left(\frac{\eta}{\eta_0}\right)^{-Q_1},\qquad
 \frac{\eta}{\etainit}\rightarrow+\infty
\end{equation}
and
\begin{equation}\label{ksmall}
 k\rightarrow k_3\left(\frac{\eta}{\eta_0}\right)^{-Q_3},\qquad
 \frac{\eta}{\etainit}\rightarrow0.
\end{equation}
We shall thus discuss the dynamics of the perturbations when their mode is
aligned along a principal axis (\S~\ref{4b}) and their
contribution to $\Xi^{(2)}_\lambda(\bf k,\eta)$  (\S~\ref{4c}). These two
steps can be performed completely analytically so that we can then discuss
the general behaviour of $\langle \Xi^{(2)}({\bf x},\eta) \rangle$
at late time (\S~\ref{4d}) and early time (\S~\ref{4e}).

\subsection{Dynamics of the perturbation for modes aligned along a principal axis}\label{4b}

\subsubsection{General behaviour}

We assume that only one of the components of the Fourier space,
indexed by $I$, satisfies $k_I\not=0$. It is then clear that
$k^2 = k_I^2(\eta/\eta_0)^{-2Q_I}$ so that
\begin{equation}\label{e411}
 \Spar = 2 Q_I.
\end{equation}
Then, since $\Spar'=0$, we conclude from Eq.~(\ref{background_sigma})
that $\sum_a \Sva^2=0$ because of Eq.~(\ref{background_sigma2}) and thus
\begin{equation}\label{svagen}
 \Sva=0.
\end{equation}
This implies, from Eq.~(\ref{eqphia}), that 
\begin{equation}
\Phi_a=0.
\end{equation}
It follows from Eq.~(\ref{background_sigma3}) that  $\Stl'=0$, and it can be checked that 
\begin{equation}\label{e414}
  \left\vert\Stplus\right\vert =\frac{3}{\sqrt{2}}\Delta_I,
  \qquad
 \Stcross = 0
 \end{equation}
where we have defined
\begin{equation}
 \Delta_I \equiv  q_{{\rm max}(i\not=I)}-q_{{\rm min}(i\not=I)} 
\end{equation}
which is explicitely given by
\begin{eqnarray}
 \Delta_I =\frac{2}{\sqrt{3}}\left\vert\cos\varpi_I\right\vert=\sqrt{\frac{4}{3}-3 q_I^2}
 =2\sqrt{\frac{(1-Q_I^2)}{3}}.
\end{eqnarray}
The scalar perturbations are then given algebraically by
\begin{equation}\label{e:temp2gen}
 X = \frac{\Stplus}{2-3q_I} E_{+}=\frac{\Stplus}{2(1-Q_I)} E_{+}
\end{equation}
that derives from Eq.~(\ref{e:temp2}) from which we deduce
that Eqs.~(\ref{magic_equation}-\ref{e:scal2b}) reduce to
\begin{eqnarray}\label{phipsigen}
 k^2\Psi &=& -\HH X',\qquad
  k^2\Phi  =-(1-2Q_I)\HH X'.
\end{eqnarray}

\begin{figure}
\includegraphics[width=8.5cm]{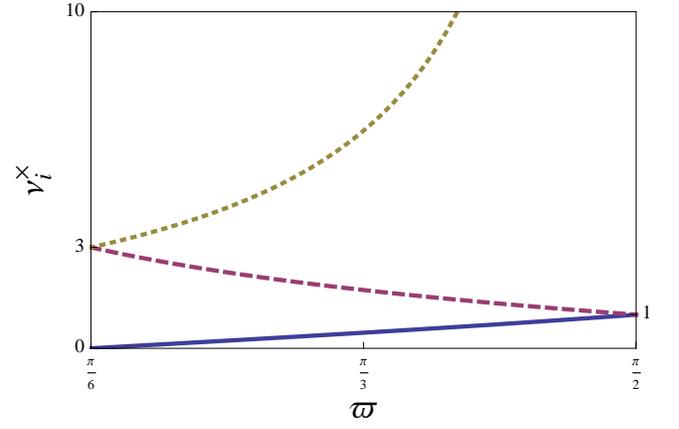}
\caption{Evolution of the index of the Bessel function, $\nu^\times_i$, with $\varpi$ ($\nu^\times_1$ in solid line,
$\nu^\times_2$ in dashed line and $\nu^\times_3$ in dotted line). We recall that
$\nu^+_i=0$ whatever $\varpi$.}
 \label{f2}
\end{figure}

\begin{figure*}
\includegraphics[width=6cm]{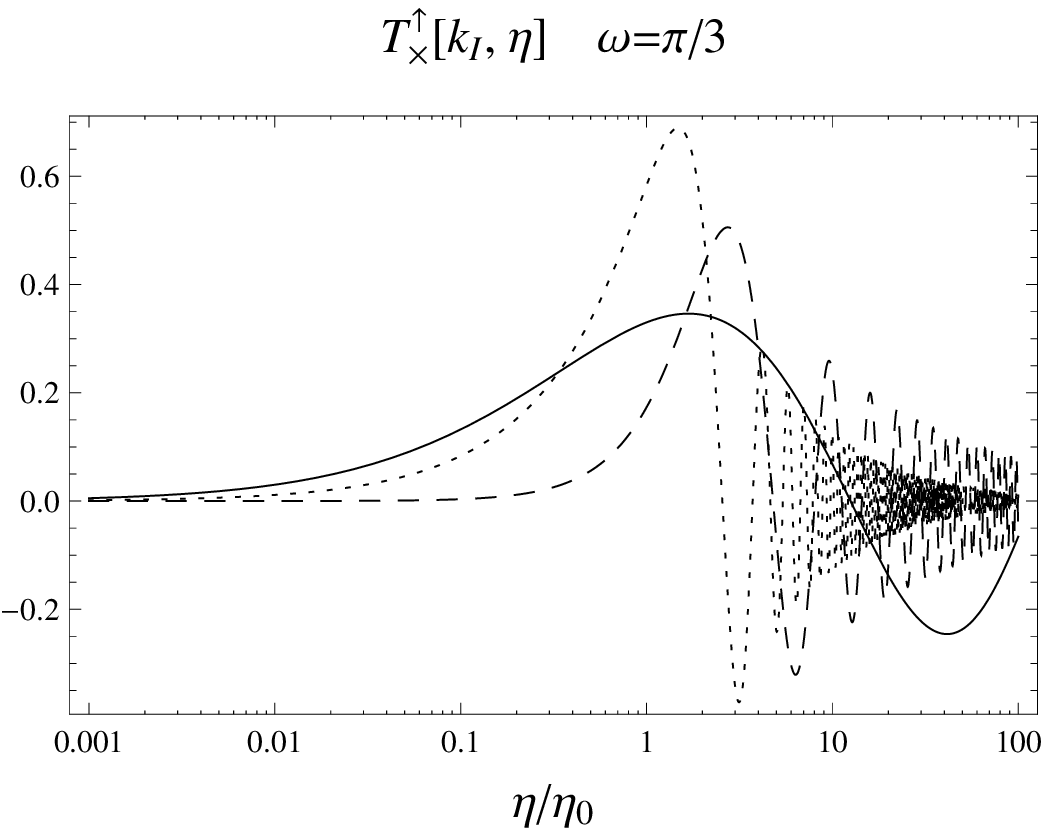}\includegraphics[width=6cm]{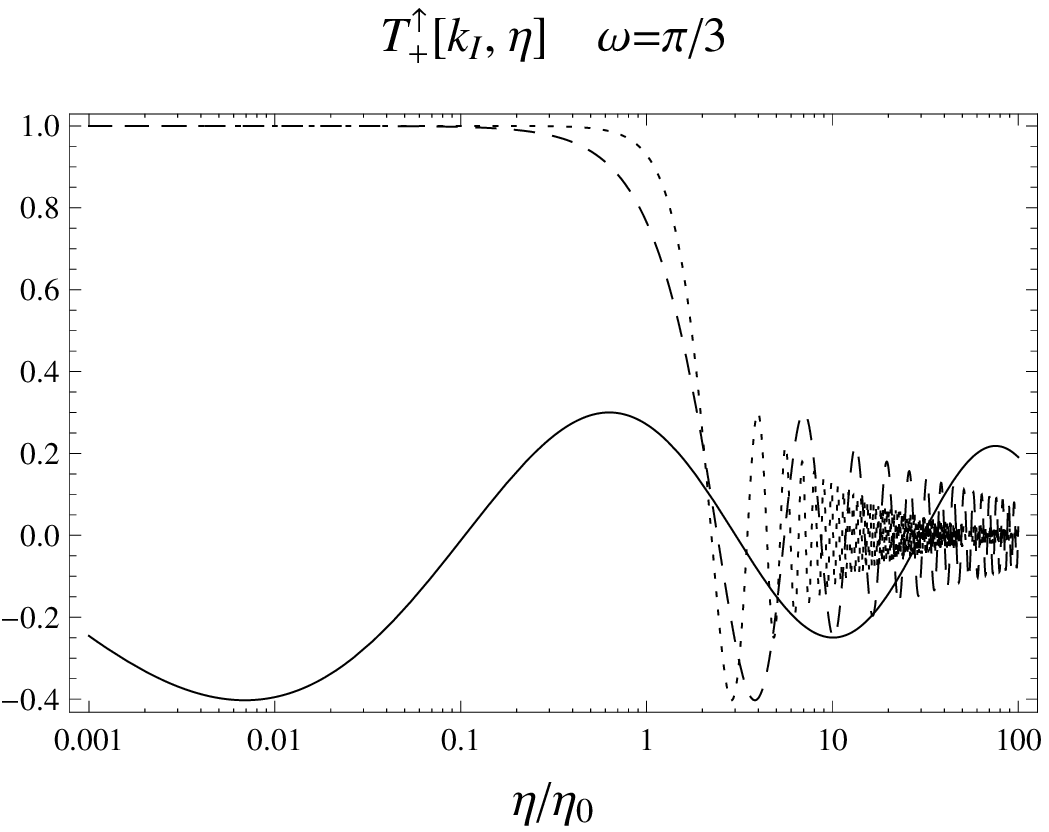}\includegraphics[width=6cm]{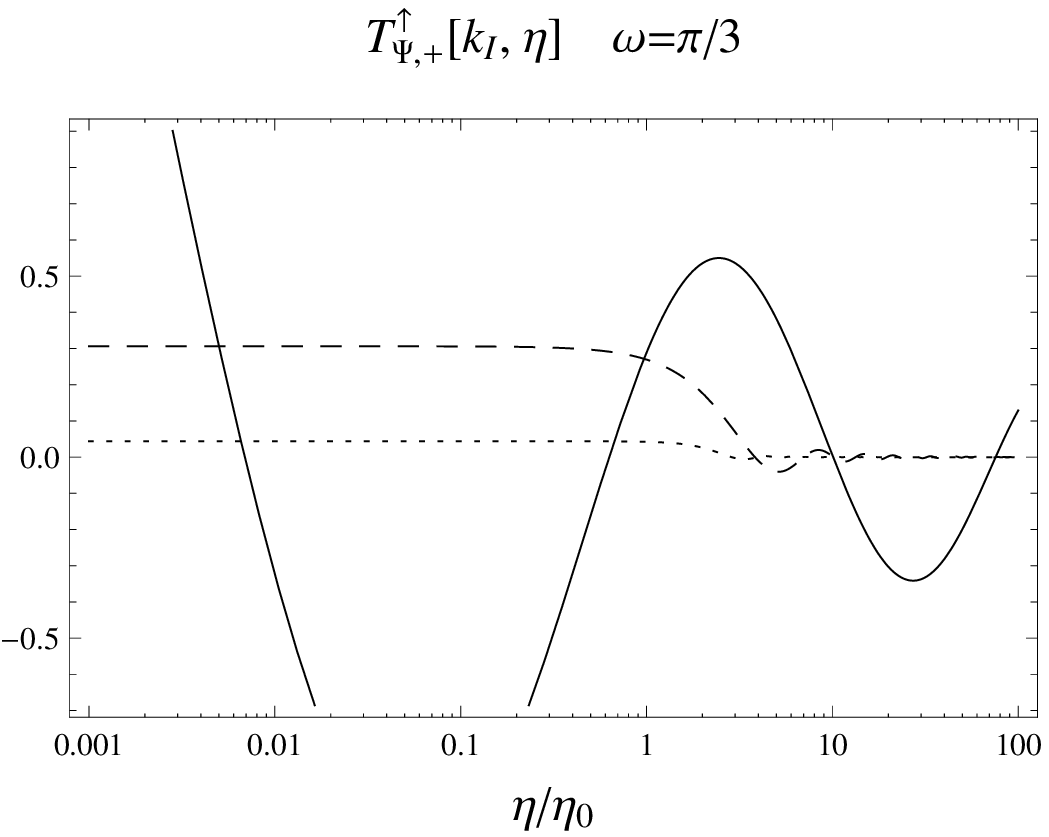}
\caption{Evolution of $T_\times^{\uparrow}$ (left), $T_+^{\uparrow}$ (middle)
and $T_{\Psi,+}^{\uparrow}$ (right) for a Kasner universe with $\varpi=\frac{\pi}{3}$,
i.e. with $\lbrace q_i\rbrace=(-1/\sqrt{3},0,1/\sqrt{3})$) for the
modes aligned with the axis 1 (dotted), 2 (dashed) and 3 (solid),
assuming that $k_I \eta_0=1$.} 
\label{fCP1in}
\end{figure*}

The only equations to solve are Eqs.~(\ref{eqformu}) for the gravity. They decouple
for each polarisation and lead to the system
\begin{eqnarray}
 && \mu_+''+\left[k_I^2 \left(\frac{\eta}{\eta_0}\right)^{-2Q_I} +\frac{1}{4\eta^2}
  \right]\mu_+
 =0,\label{evomualigned1}\\
  &&\mu_\times''+
  \left[k_I^2\left(\frac{\eta}{\eta_0}\right)^{-2 Q_I} +\frac{1-9\Delta_I^2}{4\eta^2}
  \right]  
  \mu_\times=0.\label{evomualigned2}
\end{eqnarray}
These two equations are compatible with the general equations~(\ref{mu2axial}-\ref{omega2axial2})
obtained for $\varpi=\frac{\pi}{6}$ and, in this particular case, with those derived in Ref.~\cite{emir}.
The general solutions of Eqs.~(\ref{evomualigned1}-\ref{evomualigned2}) can both be obtained as 
[see Eqs.~(\ref{trucjp1}-\ref{trucjp2})] the linear combination~(\ref{e338})
with 
\begin{equation}
 T^{\uparrow}_{\lambda}(k_I,\eta)={\cal J}^\lambda_{\nu_I}(k_I,\eta),\quad
 T^{\downarrow}_{\lambda}(k_I,\eta)={\cal N}^\lambda_{\nu_I}(k_I,\eta)
\end{equation}   
where
\begin{eqnarray}
  {\cal J}^\lambda_{I}(k_I,\eta)&=& J_{\nu_I^\lambda}
   \left[\frac{k_I\eta_0}{1-Q_I}\left(\frac{\eta}{\eta_0}\right)^{1-Q_I} \right],\\
  {\cal N}^\lambda_{I}(k_I,\eta)&=& N_{\nu_I^\lambda}
   \left[\frac{k_I\eta_0}{1-Q_I}\left(\frac{\eta}{\eta_0}\right)^{1-Q_I} \right].
\end{eqnarray}
Here, $J_{\nu_I^\lambda}$ and $N_{\nu_I^\lambda}$ are the Bessel and Newmann functions of index 
\begin{equation}
 \nu_I^\times \equiv
 \frac{3 \Delta_I}{2(1-Q_I)}=\sqrt{\frac{3(1+Q_I)}{(1-Q_I)}}\,,\qquad \nu_I^+=0,
\end{equation}
and $\nu_I^\times$ is depicted on Fig.~\ref{f2}
(note that $\nu_I^\times$ is always non-negative while
$\Stplus$ can be negative). It is then clear that
\begin{equation}\label{xki}
 |X| = \nu_I^\times \frac{E_+}{\sqrt{2}}
\end{equation}
so that the two Bardeen potentials are given by
\begin{eqnarray}
 \hspace{-0.2cm}\Psi &=& -\frac{1}{k_I^2\eta_0^2}
 \left(\frac{\eta}{\eta_0}\right)^{-2(1-Q_I)}\frac{\Stplus}{4(1- Q_I)}
 \frac{\partial E_+}{\partial \ln \frac{\eta}{\eta_0}} \label{bb1} \\
 \hspace{-0.2cm}\Phi &=& (1-2 Q_I)\Psi.\label{bb2}
\end{eqnarray}
If follows that for these particular modes, the solution of the dynamics is obtained
completely analytically. Figures~\ref{fCP1in} give an example
of the behaviour of the tensor and scalar modes.

\subsubsection{Late time behaviour}

The late time behaviour of the perturbations can be easily understood using the
large argument expansions~(\ref{besselJinf}) and~(\ref{besselNinf}) of the
Bessel functions, since
\begin{eqnarray}
 T^{\uparrow}_{\lambda}&\simeq&\sqrt{\frac{2}{\pi\ell_1}}\left(\frac{\eta}{\eta_0}\right)^{-\frac{1-Q_1}{2}}
\cos\left[\ell_1\left(\frac{\eta}{\eta_0}\right)^{1-Q_1}
 \!\!\!\!\!\!-\nu_1\frac{\pi}{2} -\frac{\pi}{4} \right]  
\nonumber\\
 T^{\downarrow}_{\lambda}&\simeq&\sqrt{\frac{2}{\pi\ell_1}}\left(\frac{\eta}{\eta_0}\right)^{-\frac{1-Q_1}{2}}
\sin\left[\ell_1\left(\frac{\eta}{\eta_0}\right)^{1-Q_1}
 \!\!\!\!\!\!-\nu_1\frac{\pi}{2} -\frac{\pi}{4} \right]  
 \nonumber
\end{eqnarray}
with
\begin{equation}
 \ell_1= \frac{2k_1\eta_0}{2-3q_1}=\frac{k_1\eta_0}{1-Q_1}\,.
\end{equation}
It follows that the late time behaviour is in general of the form
\begin{equation}\label{EqAsymptotEinfty}
 E_\lambda \sim\sqrt{\frac{2}{\pi\ell_1}}\left(\frac{\eta}{\eta_0}\right)^{-\frac{1-Q_1}{2}}
 A^{(\infty)}_\lambda\cos\left[\ell_1\left(\frac{\eta}{\eta_0}\right)^{1-Q_1}
  \!\!\!\!\!\!+\varphi^{(\infty)}_\lambda \right],
\end{equation}
where $A^{(\infty)}_\lambda$ and $\varphi^{(\infty)}_\lambda$ are two constants that can easily be obtained
in terms of $a_\lambda$ and $b_\lambda$.
$X$ is then proportional to $E_+$ [see Eq.~(\ref{xki})] and $\Phi_a=0$, which completely determines
the solution [see Eqs.~(\ref{bb1}-\ref{bb2})].

\subsubsection{Early time behaviour}\label{secgen3}

Asymptotically, the dominant term of the wavevector is aligned along
the direction expanding the fastest, i.e $i=3$. The two polarisations
behave differently because $\Delta_3\not=0$. Using the small argument 
expansions~(\ref{besselJ0}) and~(\ref{besselN0b}) of the Bessel function, 
we obtain that, since $1-Q_3\geq0$,  when $\eta\rightarrow0$
\begin{eqnarray}
 T^\uparrow_\times &\sim& \frac{1}{\Gamma(1+\nu^\times_3)}
  \left[\frac{\ell_3}{2}\left(\frac{\eta}{\eta_0}\right)^{1-Q_3} \right]^{\nu_3^\times} \nonumber\\
 T^\downarrow_\times &\sim&  \frac{\Gamma(\nu_3^\times)}{\pi}
  \left[\frac{\ell_3}{2}\left(\frac{\eta}{\eta_0}\right)^{1-Q_3} \right]^{-\nu_3^\times}
  \label{e430}
 \end{eqnarray}
with 
\begin{equation}
 \ell_3=\frac{k_3\eta_0}{1-Q_3}\,.
\end{equation} 
These expressions hold as long as $Q_3\not=1$, that is for $\varpi\not=\frac{\pi}{2}$, a
particular case that is discussed in Appendix~\ref{secpi2}. The transfer functions for the $+$ polarisation 
behave as
\begin{eqnarray}
 T^\uparrow_+ &\sim& \left[1-\left(\frac{\ell_3}{2}\right)^2\left(\frac{\eta}{\eta_0}\right)^{2(1-Q_3)} \right]\\
 T^\downarrow_+ &\sim& \frac{2}{\pi}\left[\ln\left(\frac{\ell_3}{2} \left(\frac{\eta}{\eta_0}\right)^{1-Q_3}\right)
 +\gamma_E\right].
 \label{e432}
 \end{eqnarray}

\subsection{Weyl tensor evolution for modes aligned along a principal axis}\label{4c}

For these modes, the expressions of $\Xi^{(1)}_\lambda(\bf k,\eta)$ and
$\Xi^{(2)}_\lambda(\bf k,\eta)$ turn out to simplify greatly and can be expressed
in terms of the dimensionless parameter
\begin{equation}\label{defx}
 x \equiv k \eta = x_I \eta\,.
\end{equation}

First,  $\Xi^{(1)}_\lambda(\bf k,\eta)$  can be expressed in terms of the two transfer functions
and their first derivatives as
\begin{equation}\label{EqC2o1general}
\Xi^{(1)}(k_I,\eta)=\frac{1}{2(1-Q_I)(1+2 Q_I) x^2}\sum_{i=0,1} f_+^{(i)} \frac{\partial^i T_+}{\partial (\ln \eta)^i}\,.\nonumber
\end{equation}
The two coefficients $f_+^{(i)}$ can be computed and are given by
\begin{eqnarray}
f_+^{(0)} &=&\sqrt{2} \Delta_I[ -6 (1+2 Q_I) x^2 + 2 x^4],\\
f_+^{(1)} &=&\sqrt{2} \Delta_I[ -9 (1+2 Q_I)  + (7+ 2 Q_I) x^2].
\end{eqnarray}
Interestingly, the $\times$ polarisation does not contribute. In the
above expression $T_+$ stands either for $T_+^{\downarrow}$ 
or $T_+^{\uparrow}$ depending on the solution (decaying, growing, or any linear combination
of them) that is considered.
When $\Delta_I=0$, $\Xi^{(1)}$ vanishes identically and, again, since $\langle \Xi^{(1)} \rangle = 0$ 
we shall not consider this linear part any further.

The second order $\Xi^{(2)}$ can be expanded as
\begin{eqnarray}
\Xi^{(2)}_\lambda(k_I,\eta)&=&\frac{1}{6(1+2 Q_I)^2(1-Q_I)}\label{expchi2}\\
&&\times \sum_{i=0,1}\sum_{j=0,1}
f_{\lambda}^{(i,j)}\frac{\partial^i T_\lambda}{\partial (\ln \eta)^i}
\frac{\partial^j T_\lambda}{\partial (\ln \eta)^j}\nonumber
\end{eqnarray}
with the same conventions as for $\Xi^{(1)}$. 
The eight coefficients $f_\lambda^{(i,j)}$ are explicitely given by
\begin{eqnarray}
f_{+}^{(0,0)}&=&3 (1 + 2 Q_I)^2 (13 + 8 Q_I) (1 + Q_I) \nonumber\\
&&\qquad+ 6 x^2 (-1 + 4 Q_I^2) + 8 x^4\\
f_{\times}^{(0,0)}&=&6(1+2 Q_I)^2(1-Q_I^2)+18 x^2+ 8 x^4,
\end{eqnarray}
\begin{eqnarray}
f_{+}^{(0,1)}&=&f_{+}^{(1,0)}\nonumber\\
  &=&\frac{1}{x^2}\left[3(1 + 2 Q_I)^2 (1 +  Q_I)(22+25
  Q_I-2 Q_I^2)\right. \nonumber\\
&&+2(1+2 Q_I)(-26-29 Q_I+ Q_I^2)x^2 \nonumber\\
&&\left. +4(1-Q_I)x^4  \right]\\
f_{\times}^{(0,1)}&=&f_{\times}^{(1,0)}\nonumber\\
  &=&-4(1-Q_I)[(1+2 Q_I)^2-x^2]
\end{eqnarray}
and
\begin{eqnarray}
f_{+}^{(1,1)}&=&x^{-4}\left[9 (1 + 2 Q_I)^2(1 + Q_I)(3+ 2 Q_I+4 Q_I^2)\right.\nonumber\\
&&\qquad\times(5-2 Q_I) +2 (13 + 4 Q_I-8 Q_I^2)x^4+8 x^6\nonumber\\
&& \left. +9 (1 + 2 Q_I)(1 + Q_I)(-17-18 Q_I+8 Q_I^2)x^2\right]\nonumber\\
f_{\times}^{(1,1)}&=&2[(1+2 Q_I)^2+ 4x^2].
\end{eqnarray}

\subsection{Asymptotic behaviour at late times}\label{4d}

At late time ($\eta \to \infty$), only the growing mode dominates and we can safely
neglect the decaying mode since it would only lead to a redefinition
of the phase $\varphi_\infty$; see Eq.~(\ref{EqAsymptotEinfty}).
As can be concluded from the evolution~(\ref{klarge}) of the wave number,
the parameter $x$ defined in Eq.~(\ref{defx}) behaves as
\begin{equation}
 x \to k_1 \eta_0 \left(\frac{\eta}{\eta_0}\right)^{1-Q_1}
\end{equation}
which is always a growing function of $\eta$ (remember $q_1<0$ and
thus $Q_1<0$; see Fig.~\ref{f1}). 
We thus need to take the
limit when $x\to \infty$ of the transfer functions $T^{\uparrow}_\lambda$
and of their derivatives. From Eq.~(\ref{EqAsymptotEinfty}), we conclude
that
\begin{equation}
 T^{\uparrow}_\lambda \simeq \sqrt{\frac{2(1-Q_1)}{\pi x}}\cos\left[\frac{x}{1-Q_1}\right].
\end{equation}
Since
\begin{equation}
 \frac{\partial x}{\partial \ln \eta}=\left(1-Q_1\right)x,
\end{equation}
it implies that, at leading order in $x$,
\begin{eqnarray}
\frac{\partial T^{\uparrow}_\lambda}{\partial \ln \eta} \simeq
 - \sqrt{\frac{2 x (1- Q_1)}{\pi}} \sin\left[\frac{x}{1-Q_1}\right].
\end{eqnarray} 
Using the expansion~(\ref{expchi2}), we conclude that
\begin{eqnarray}\label{LimXiLate}
\Xi^{(2)}_\lambda({\bf k},\eta) &\to& \frac{4x^2}{3} \frac{\left[\left({\frac{\partial T^{\uparrow}_\lambda}{\partial \ln \eta}}\right)^2+ x^2 {T^{\uparrow}_\lambda}^2\right]}{(1+2 Q_1)^2(1-Q_1)} \nonumber\\
&\to&\frac{8 x^3}{3 \pi (1+2 Q_1)^2}\,.\label{e444}
\end{eqnarray}
This expression is valid for any Kasner spacetime but for the particular case $\varpi=\frac{\pi}{2}$
that is discussed in Appendix~\ref{secpi2} (since in that case $Q_1=-\frac{1}{2}$).

It follows that
\begin{eqnarray}
\Xi^{(2)}_\lambda({\bf k},\eta) \propto x^3 \propto \eta^{3(1-Q_1)},
\end{eqnarray}
which is unbounded at late time.

\subsection{Asymptotic behaviour at early times}\label{4e}

As can be concluded from the evolution~(\ref{ksmall}) of the wave number,
the parameter $x$ defined in Eq.~(\ref{defx}) behaves, when $\eta \to 0$, as
\begin{equation}
 x \to k_3 \eta_0 \left(\frac{\eta}{\eta_0}\right)^{1-Q_3}
\end{equation}
which is always a decreasing function of $\eta$ (except if
$\varpi=\frac{\pi}{2}$, see Appendix~\ref{secpi2} for that particular case).
It follows that we need to evaluate the limit $x\to 0$ of the transfer
functions.

At early time, one can however not disregard the decaying mode since it
diverges faster when $\eta\rightarrow0$. 
The expression~(\ref{expchi2}) behaves as
\begin{eqnarray}\label{EqlimC2}
&&\Xi^{(2)}({\bf k},\eta) \to\frac{(1+ Q_3)}{2(1- Q_3)} (13 + 8 Q_3) T_+ T_+\\
&&+\frac{ (1 + Q_3)}{ (1-Q_3) x^2}(22+25
  Q_3-2 Q_3^2) T_+ \frac{\partial T_+}{\partial \ln \eta}\nonumber\\
&&+\frac{3 (1 +  Q_3)}{2(1- Q_3) x^4}(5-2 Q_I) (3+ 2Q_3+4 Q_3^2) \left(\frac{\partial T_+}{\partial \ln \eta}\right)^2\nonumber\\
&&+(1+Q_3) T_\times T_\times-\frac{4}{3} T_\times \frac{\partial T_\times}{\partial \ln \eta}+\frac{1}{3(1-Q_3)}\left(\frac{\partial T_\times}{\partial \ln \eta}\right)^2. \nonumber
\end{eqnarray}
In order to determine the behaviour of the transfer functions, we consider two cases.

First, if we consider only the growing mode [i.e. $a_\lambda \neq 0$ and $ b_\lambda=0$
in Eq.~(\ref{e338})] then we deduce from the expressions~(\ref{e430}) and~(\ref{e432})
of the transfer functions that
\begin{eqnarray}
{\frac{\partial T^{\uparrow}_\times}{\partial \ln \eta}}&\simeq&
\frac{3}{2}\Delta_3 T^{\uparrow}_\times\,,\nonumber\\
{\frac{\partial T^{\uparrow}_+}{\partial \ln
    \eta}}&\simeq&-\frac{x^2}{2(1-Q_3)} T^{\uparrow}_+.
\end{eqnarray}
At leading order, the contributions of the two polarisations to $\Xi^{(2)}$ are thus
\begin{eqnarray}
\Xi^{(2)}_+(k_I,\eta) &\to& \frac{9(1-4 Q_3)^2(1+ Q_3)}{8(1- Q_3)^3}{T_+^{\uparrow}}^2\,, \nonumber\\
\Xi^{(2)}_\times(k_I,\eta)&\to&2 (1+Q_3-\Delta_3) {T^{\uparrow}_\times}^2.
\end{eqnarray}
Eqs.~(\ref{e430}) and~(\ref{e432}) tells that $T^{\uparrow}_\times\sim x^{\nu_3^\times}$
and $T^{\uparrow}_+\sim x^0$ so that $\Xi^{(2)}_\times(k_I,\eta) \to0$ (since
$\nu_3^\times>0$) and $\Xi^{(2)}_+(k_I,\eta)\to {\rm const.}$ 
Thus, the contribution of the growing mode leads to a bounded contribution to $\Xi^{(2)}$.

On the other hand, if we consider the decaying mode [i.e. $a_\lambda= 0$ and $ b_\lambda\not=0$
in Eq.~(\ref{e338})], i.e. the most diverging mode when $x\rightarrow0$, then
\begin{eqnarray}
{\frac{\partial T^{\downarrow}_\times}{\partial \ln \eta}}&\simeq&-\frac{3}{2}\Delta_3 T^{\downarrow}_\times\nonumber\\
{\frac{\partial T^{\downarrow}_+}{\partial \ln \eta}}&\simeq&\frac{2(1-
Q_3)}{\pi}\gg x^2T^{\downarrow}_+.
\end{eqnarray}
We conclude that at leading order the contributions of the two polarisations
to $\Xi^{(2)}$ are
\begin{eqnarray}
\Xi^{(2)}_+({\bf k},\eta) &\to& \frac{6 (1-Q_3^2)}{\pi^2 x^4} (5-2 Q_3) (3+2 Q_3+4 Q_3^2) \nonumber\\
\Xi^{(2)}_\times({\bf k},\eta)&\to&2(1+Q_3+ \Delta_3) {T^{\downarrow}_\times}^2.
\end{eqnarray}
Since, from Eqs.~(\ref{e430}) and~(\ref{e432}), $T^{\downarrow}_\times \propto x^{-\nu_3^\times}$ with
$\nu_3^\times \ge 3$ for all cases, the $\times$ contribution diverges faster when $\eta \to 0$ and
we conclude that
\begin{eqnarray}
\Xi^{(2)}({\bf k},\eta)&\propto& x^{-2\nu_3^\times}\propto \eta^{-3\Delta_3}.
\end{eqnarray}

\subsection{Discussion}

During the cosmic evolution, for any modes, the components of the shear
evolve from an early value corresponding to the situation of the mode
being aligned to the third spatial axis (index $I=3$) to a situation corresponding to the mode
being aligned to the first spatial axis (index $I=1$). This dynamics arises from the fact that the
Euler angles are not constant over time and that $\beta$ evolves from 0 to $\frac{\pi}{2}$
and $\gamma$ from $\frac{\pi}{2}$ to 0 (except in the axisymmetric
cases $\varpi=\frac{\pi}{6},\frac{\pi}{2}$ for which the value of $\gamma$ does not matter).
This implies that $\Spar$ varies from $3q_3$ to $3q_1$, $\Sva$ and $\Stcross$ from 0 to 0 but
with a transient deviation from 0 and $\Stplus$ from $\frac{3}{\sqrt{2}}\Delta_3$
to $\frac{3}{\sqrt{2}}\Delta_1$, as depicted on Fig.~\ref{f3}.

It follows that the asymptotic behaviour of the perturbation of the Weyl tensor
can be discussed by looking at these two particular regimes. Figure~\ref{fcp1axial}
shows how a numerical solution interpolates between these two regimes.

Using this procedure, analog to the one developed in Ref.~\cite{emir}, we conclude that
the Kasner spacetime is unstable at late time since $\delta^{(2)} C^2$ decreases with
time slower than $\bar C^2$, which means that the Weyl tensor of the perturbation
always dominate the Weyl tensor of the homogeneous space at late time. Indeed, for a given 
mode the late time asymptotic regime is reached when
\begin{equation}
 \eta \gg \eta^{\rm late}_{\bf k}=\hbox{max}\left[\left(\frac{k_2}{k_1}\right)^{\frac{1}{(Q_2-Q_1)}},
 \left(\frac{k_3}{k_1}\right)^{\frac{1}{(Q_3-Q_1)}}\right]\,.
\end{equation}
At any given time $\eta$, we denote by ${\cal B}({\bf k},\eta)$ the set of modes such that
$\eta > \eta^{\rm late}_{\bf k}$ so that Eq.~(\ref{e47}) implies that
\begin{equation}\label{e47b}
\langle \Xi^{(2)}({\bf x},\eta) \rangle > \int_{{\cal B}({\bf k},\eta)}
\frac{\dd^3 {\bf k} }{(2 \pi)^3} P_{\rm init}(k_i) \Xi^{(2)}(\bf k,\eta)\,,
\end{equation}
since $\Xi^{(2)}({\bf k},\eta)$ is positive. Given that $\Xi^{(2)}(\bf k,\eta)$ is converging toward
$\Xi^{(2)}(k_1,\eta)$ which is diverging at late times, we deduce that $\Xi^{(2)}(\bf k,\eta)$ is diverging at late times.
This is compatible with the analysis of Ref.~\cite{emir} for the case of an axisymmetric
Kasner spacetime, as well as with earlier findings~\cite{ppu2,emir} on the amplification
of gravity waves in anisotropic inflation during the shear dominated era, which
can be described by a Kasner era of finite duration.

At early time the conclusion depends on the choice of the modes that
are considered. We have shown that if decaying modes are present
then the spacetime is unstable when $\eta\to0$, following the same argument
as above but with the modes such that
\begin{equation}
 \eta \ll \eta^{\rm early}_{\bf k}=\hbox{min}\left[\left(\frac{k_2}{k_3}\right)^{\frac{1}{(Q_3-Q_2)}},
 \left(\frac{k_1}{k_3}\right)^{\frac{1}{(Q_3-Q_1)}}\right],
\end{equation}  
whereas it is stable if one imposes to have only growing modes since then $\Xi^{(2)}$
remains bounded. As pointed out in~\cite{emir} (see Appendix~B),
this result is compatible with the analysis by Ref.~\cite{bkl1} where
it was pointed out that a condition on the perturbations should
be imposed for the spacetime to be stable when $\eta\to0$. As explained
in Ref.~\cite{emir}, this condition precisely kills the decaying modes
${\cal N}_I^\lambda$ for both perturbations.

\section{Axially symmetric case with $\varpi=\frac{\pi}{6}$}\label{secaxial}

In order to compare our analysis to the existing literature, we consider the
particular case of an axially symmetric case with $\varpi=\frac{\pi}{6}$. 
This particular case was studied in Ref.~\cite{emir} and corresponds to
the axisymmetric case
\begin{equation}
q_1=-\frac{2}{3},\qquad q_2=q_3=\frac{1}{3}
\end{equation}
so that
\begin{equation}
\Delta_1=0,\qquad \Delta_2=\Delta_3=1.
\end{equation}
In that case, and in that case only, we perform a permutation 
of the axis ($3\to 1$, $2 \to 3$ and $1 \to 2$) with respect to the
parameterisation of angles given in \S~\ref{SecModeDecomposiiton}
so that the special direction is the first axis (labelled by the index
$I=1$).

After deriving the gravity wave equation (\S~\ref{5a}) and showing
that the two polarisations are decoupled at all times (and not
only asymptotically as for a generic case), and expressing
the vector and scalar perturbation (\S~\ref{5b}), we investigate
the late (\S~\ref{5c}) and early (\S~\ref{5d}) time behaviours 
(and the interpolation of the exact numerical integration
between these two asympotic regimes in \S~\ref{5e}).

\subsection{Gravity wave equation}\label{5a}

\begin{figure}
 \includegraphics[width=8.5cm]{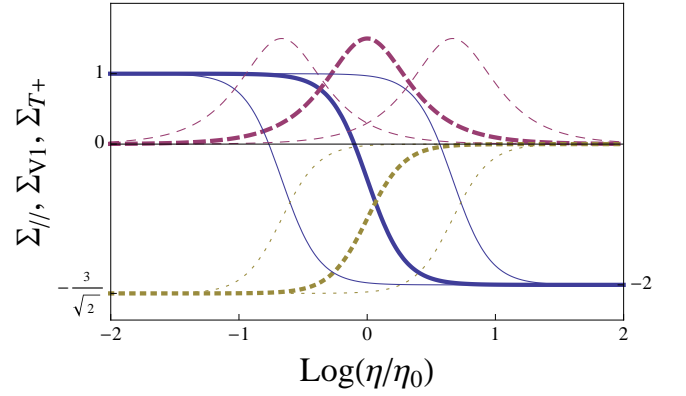}
 \caption{Evolution of $\Spar$ (solid), $\Svun$ (dashed)
 and $\Stplus$ (dotted) for $k_1\eta_0=k_\perp\eta_0=1$ (heavy)
 and  $k_1\eta_0=10, k_\perp\eta_0=1$ (light) or
  $k_1\eta_0=1, k_\perp\eta_0=10$ (light)
 for a Kasner universe with $\varpi=\frac{\pi}{6}$.}\label{f3}
\end{figure}

We set $k_\perp^2=k_2^2+k_3^2$, so that Eq.~(\ref{deff2b}) takes the form
\begin{equation}
 k^2=k_1^2\left(\frac{\eta}{\eta_0}\right)^2 + k_\perp^2\left(\frac{\eta}{\eta_0}\right)^{-1}
\end{equation}
and we shall assume that $k_1k_\perp\not=0$, since otherwise
we are back to \S~\ref{4b}.
Because of the rotational invariance, Eq.~(\ref{euler2})
implies that the Euler angle $\gamma$ is constant
\begin{equation}
 \tan\gamma=\frac{k_2}{k_3},
\end{equation}
so that Eq.~(\ref{euler3}) implies that $\alpha'=0$.
We can thus choose
\begin{equation}
 \alpha=0.
\end{equation}
The Euler angle $\beta$ is obtained from Eq.~(\ref{euler1})
\begin{equation}
 \cos\beta = \frac{k_1}{k}\left(\frac{\eta}{\eta_0}\right),
 \quad
 \sin\beta = \frac{k_\perp}{k}\left(\frac{\eta}{\eta_0}\right)^{-\frac12}.
\end{equation}
It follows that Eqs.~(\ref{euler4}-\ref{euler5}) simplify to
\begin{equation}
 \omega_1^i=\left(\begin{array}{c}\cos\gamma\cos\beta\\
                             \cos\beta\sin\gamma\\
                             -\sin\beta \end{array}\right),\qquad
 \omega_2^i =\left(\begin{array}{c}-\sin\gamma\\
                             \cos\gamma\\
                             0\end{array}\right).
\end{equation}
The components of the shear then take very simple
expressions. Eq.~(\ref{s1}) reduces to
\begin{equation}\label{s1axial}
 \Spar =\sin^2\beta-2 \cos^2\beta,
\end{equation}
while Eqs.~(\ref{s2}-\ref{s3}) give
\begin{eqnarray}\label{s2axial}
 \Svun =3 \sin\beta \cos\beta,\qquad
 \Svdeux=0
\end{eqnarray}
and  Eqs.~(\ref{s4}-\ref{s5}) lead to
\begin{eqnarray}\label{s3axial}
 \Stplus =- \frac{3}{\sqrt{2}}\sin^2\beta,\qquad
 \Stcross = 0.
\end{eqnarray}
These expressions satisfy indeed Eq.~(\ref{sumsurs}) and are depicted
on Fig.~\ref{f3}. We recover the behaviour~(\ref{e411}) for $\Spar$,
(\ref{svagen}) for $\Sva$ and~(\ref{e414}) for $\Stl$.

Since $\Stcross = 0$, the two polarisations decouple and
their equations of evolution~(\ref{eqformu}) simplify to
\begin{eqnarray}\label{mu2axial}
 \mu_{\times,+}''+ \omega^2_{\times,+}(k_i,\mu)\mu_{\times,+}=0
\end{eqnarray}
with
\begin{eqnarray}
 \omega^2_{\times} &=&  k^2 + \frac{1}{4\eta^2}
     \left[1 - 2\left(\Stplus^2-\Svun^2\right)\right]
     \label{omega2axial1}\\
      \omega^2_{+} 
      &=&  k^2 + \frac{1}{4\eta^2}
     \left[1 + 2\Svun^2\left(\frac{\sqrt{2}\Stplus}{2-\Spar}-1\right)^2 \right]\label{omega2axial2}
\end{eqnarray}
where we have used that $\Spar'=-2\HH\Svun^2$ and $\Stplus'=\sqrt{2}\HH\Svun^2$.
We emphasize that our expressions agree with those of Ref.~\cite{emir}
(see for instance Eq.~(40) of that reference).

\subsection{Scalar and vector perturbations}\label{5b}

The expressions of the scalar perturbations are therefore
\begin{equation}\label{axial-X}
X = \frac{\Stplus}{2-\Spar}\,  E_{+}
\end{equation}
that derives from Eq.~(\ref{e:temp2}) from which we deduce
that Eqs.~(\ref{magic_equation}-\ref{e:scal2b}) reduce to
\begin{eqnarray}\label{axial-psi}
k^2\Psi&=&\frac{\HH}{(2-\Spar)}
\left[-\Stplus E'_+\right.\\
&& \left.+ \HH\left( \sqrt{2}\Svun^2-2\frac{\Svun^2\Stplus}{2-\Spar}\right)E_+ \right]\nonumber
\end{eqnarray}
and
\begin{equation}\label{axial-Phia}
 \Phi_a = -2{\rm i}\frac{\HH}{k}\Svun
 \left(\begin{array}{c}
 \left( \sqrt{2} - \frac{\Stplus}{2-\Spar}\right)E_+ \\
 \sqrt{2}E_\times
 \end{array}\right).
\end{equation}
The absolute value of the coefficient in Eq.~(\ref{axial-X}) evolves from $\nu_3^\times/\sqrt{2}=0$ to
$\nu_1^\times/\sqrt{2}=3/\sqrt{2}$ and $\Phi_a$ vanishes both at early and late times.

\subsection{Late time behaviour}\label{5c}

At late times, the wave-number behaves as $k\sim k_1\eta/\eta_0$
so that $x\sim k_1\eta_0(\eta/\eta_0)^2$. For the two polarisations,
the growing and decaying modes entering the solution~(\ref{e338}) behave as
\begin{eqnarray}
 T^\uparrow_\lambda &\sim& J_0\left[\frac{k_1\eta_0}{2} \left(\frac{\eta}{\eta_0}\right)^{2} \right]=J_0\left(\frac{x}{2}\right)\\
 T^\downarrow_\lambda &\sim&N_0\left[\frac{k_1\eta_0}{2} \left(\frac{\eta}{\eta_0}\right)^{2} \right]=N_0\left(\frac{x}{2}\right).
 \end{eqnarray}
The two modes combine to modify the phase so that the general solution is 
\begin{eqnarray}\label{Epi6infini}
 E_\lambda &\sim& \frac{2A^{(\infty)}_\lambda}{\sqrt{\pi x}} \cos\left[\frac{x}{2}+\varphi^{(\infty)}_\lambda \right].
\end{eqnarray}
The behaviour of the scalar and vector perturbations are therefore easily obtained
since $\beta\rightarrow\frac{\pi}{2}$, i.e.
\begin{equation}
\sin\beta\sim \frac{k_\perp}{k_1}\left(\frac{\eta}{\eta_0} \right)^{-\frac32},\qquad
\cos\beta\sim 1,
\end{equation}
so that $\Sigma_\parallel\sim-2$ while $\Svun\sim 3\sin\beta$ and
$\Stplus\sim-3\sin^2\beta/\sqrt{2}$. It follows that
\begin{eqnarray}
X &\sim& \frac{-3}{4\sqrt{2}}\left(\frac{k_\perp}{k_1}\right)^2
\left(\frac{\eta}{\eta_0} \right)^{-3} E_+
\end{eqnarray}
and 
\begin{eqnarray}
 \Phi_a& \sim&  \frac{-3{\rm i}\sqrt{2}}{k_1\eta_0}
  \left(\frac{k_\perp}{k_1}\right) \left(\frac{\eta}{\eta_0} \right)^{-7/2}
\left(\begin{array}{c}
 E_+\\
 E_\times
 \end{array}\right).
\end{eqnarray}
These expressions can then be inserted in Eq.~(\ref{expchi2}) that is valid
for $\varpi=\frac{\pi}{6}$ to conclude that at late time, the polarisations
contribute in the same way to $\Xi^{(2)}$ as
\begin{eqnarray}
  \Xi^{(2)}_\lambda&\to& \frac{8 x^3}{3 \pi}\,,
\end{eqnarray}
which is equivalent to our prior result~(\ref{e444}). The result for
the $\times$ polarisation matches the results of Ref.~\cite{emir} up to
normalisation factors of the transfer functions, but it appears to be
different from the conclusions on the $+$ polarisation. 
Asymptotically, the Fourier mode is aligned with the preferred direction of the spacetime,
and it turns out to be difficult to compare our result with 
Ref.~\cite{emir} given that most expressions of their formalism 
are singular for a mode along that special direction.
We are confident that our result is correct in that limit since from
Eq.~(\ref{LimXiLate}) we find that $\bar C^2 \Xi^{(2)}_\lambda$, which is the perturbation of the
square of the Weyl, when expressed in terms of the transfer function
takes the form
\begin{equation}
\bar C^2 \Xi^{(2)}_\lambda \to \frac{8}{S^4} \left[k^2 \left({\frac{\partial T^{\uparrow}_\lambda}{\partial \eta}}\right)^2+ k^4 {T^{\uparrow}_\lambda}^2\right].
\end{equation}
This is formally the same expression as for a pure Minkowski
space-time [see (Eq.~\ref{eb5})], which is expected when one takes the
small scale limit.

\subsection{Early time behaviour}\label{5d}

At early times, the wave-number behaves as $k\sim  k_\perp(\eta/\eta_0)^{-\frac12}$ so that
$x\sim k_\perp\eta_0(\eta/\eta_0)^{\frac12}$, which goes to zero. As in \S~\ref{4e}, we cannot neglect the
decaying mode. We can use the expression of $\Xi^{(2)}$ obtained in \S~\ref{4e} since it is well-behaved
when $q_3\rightarrow\frac{1}{3}$.

If we first consider  the growing mode [i.e. $a_\lambda \neq 0$ and $ b_\lambda=0$
in Eq.~(\ref{e338})] then we deduce that
\begin{eqnarray}
 T_+^\uparrow &\sim&J_0\left[2 k_\perp\eta_0 \left(\frac{\eta}{\eta_0}\right)^{\frac{1}{2}} \right]=
 J_0(2x)\\
 T_\times^\uparrow &\sim&J_3\left[2 k_\perp\eta_0 \left(\frac{\eta}{\eta_0}\right)^{\frac{1}{2}} \right]
 =J_3(2x),
 \end{eqnarray}
from which it follows that
\begin{eqnarray}
{\frac{\partial T^{\uparrow}_\times}{\partial \ln \eta}}\simeq\frac{3}{2} T^{\uparrow}_\times
\qquad
{\frac{\partial T^{\uparrow}_+}{\partial \ln \eta}}\simeq-x^2 T^{\uparrow}_+.
\end{eqnarray}
Asymptotically, we have from Eq.~(\ref{besselJ0}) that
\begin{eqnarray}
 T_+^\uparrow \sim 1-x^2,\qquad
 T_\times^\uparrow \sim\frac{x^3}{6}\,. 
\end{eqnarray}
Since
\begin{equation}
\cos\beta\sim \frac{k_1}{k_\perp}\left(\frac{\eta}{\eta_0} \right)^{\frac32},\qquad
\sin\beta\sim 1,
\end{equation}
$\Sigma_\parallel\sim1$ while $\Svun\sim 3\cos\beta$ and
$\Stplus\sim-3/\sqrt{2}$ and it follows that
$T_{X,\times}=T_{\Phi_a,\times}=0$ and
\begin{equation}
T_{X,+} \sim
\frac{-3 T_+}{\sqrt{2}},
\quad
 T_{\Phi_a,+} \sim \frac{-3{\rm i}\sqrt{2}}{k_\perp\eta_0}\frac{k_1}{k_\perp}
 \left(\frac{\eta}{\eta_0} \right)
   \left(\begin{array}{c}
 \frac{5}{2}T_+\\
 T_\times
 \end{array}\right).
\end{equation}
With these expressions, we conclude that at leading order, the contributions of the two polarisations 
to $\Xi^{(2)}$ are 
\begin{eqnarray}
\Xi^{(2)}_+&\to& \frac{27}{2}, 
\qquad
\Xi^{(2)}_\times \to \frac{x^6}{36},
\end{eqnarray}
which both remain bounded.

On the other hand, if we consider the decaying mode [i.e. $a_\lambda= 0$ and $ b_\lambda\not=0$
in Eq.~(\ref{e338})], i.e. the most diverging mode when $x\rightarrow0$, then
\begin{eqnarray}
 T_+^\downarrow &\sim& N_0\left[2 k_\perp\eta_0 \left(\frac{\eta}{\eta_0}\right)^{\frac{1}{2}} \right]= N_0(2x)\\
 T_\times^\downarrow &\sim& N_3\left[2 k_\perp\eta_0 \left(\frac{\eta}{\eta_0}\right)^{\frac{1}{2}} \right] = N_3(2x),
 \end{eqnarray}
the asymptotic behaviours of which are 
\begin{eqnarray}
 T_+^\downarrow \sim \frac{2}{\pi}\ln \left(x\right),\qquad
 T_\times^\downarrow \sim -\frac{2}{\pi}x^{-3}.
 \end{eqnarray}
It allows to conclude that  the contributions of the two polarisations 
to $\Xi^{(2)}$ are 
\begin{eqnarray}
\Xi^{(2)}_+\to \frac{90}{\pi^2 x^4}\,, 
\qquad
\Xi^{(2)}_\times \to \frac{20}{\pi^2 x^6}\, .
\end{eqnarray}
As expected both terms diverge when $\eta\rightarrow0$.

\subsection{Interpolation between the two regimes}\label{5e}

The transition between the two asympotic regimes is illustrated on Fig.~\ref{f3}
for the components of the shear. While the transition is quite sharp,
it is shifted depending on the value of $k_\perp/k_1$.

\begin{figure*}
\includegraphics[width=6cm]{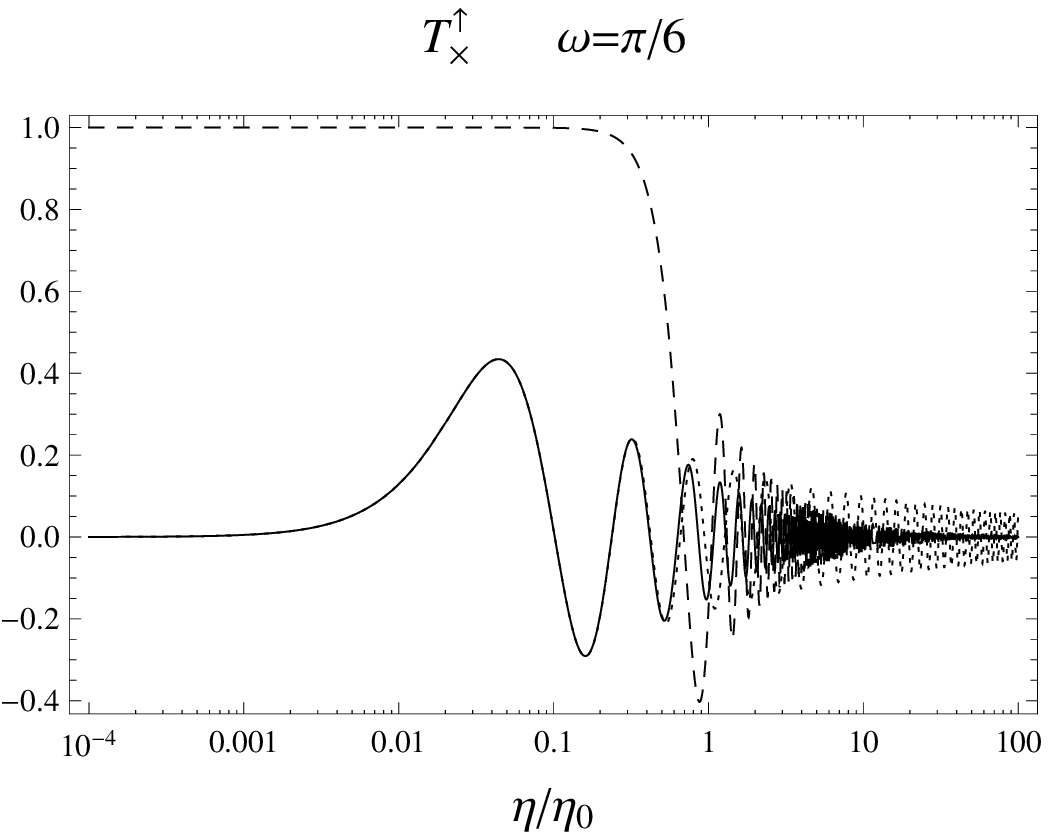}\includegraphics[width=6cm]{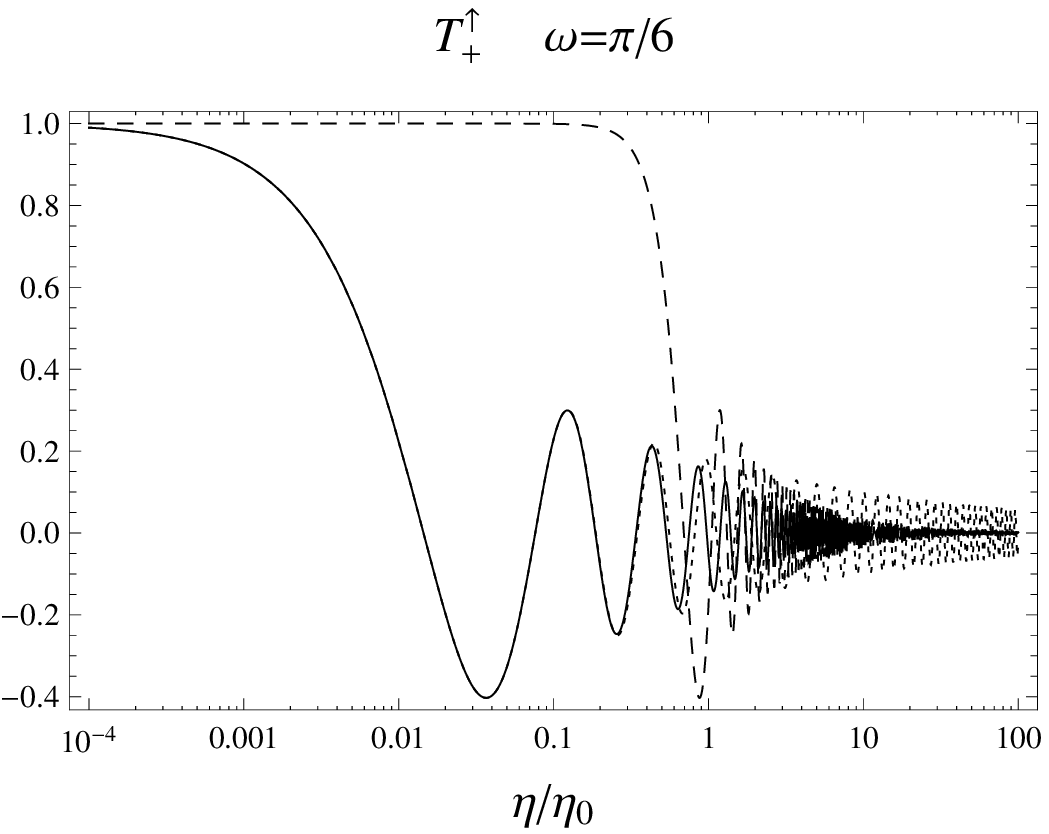}\includegraphics[width=6cm]{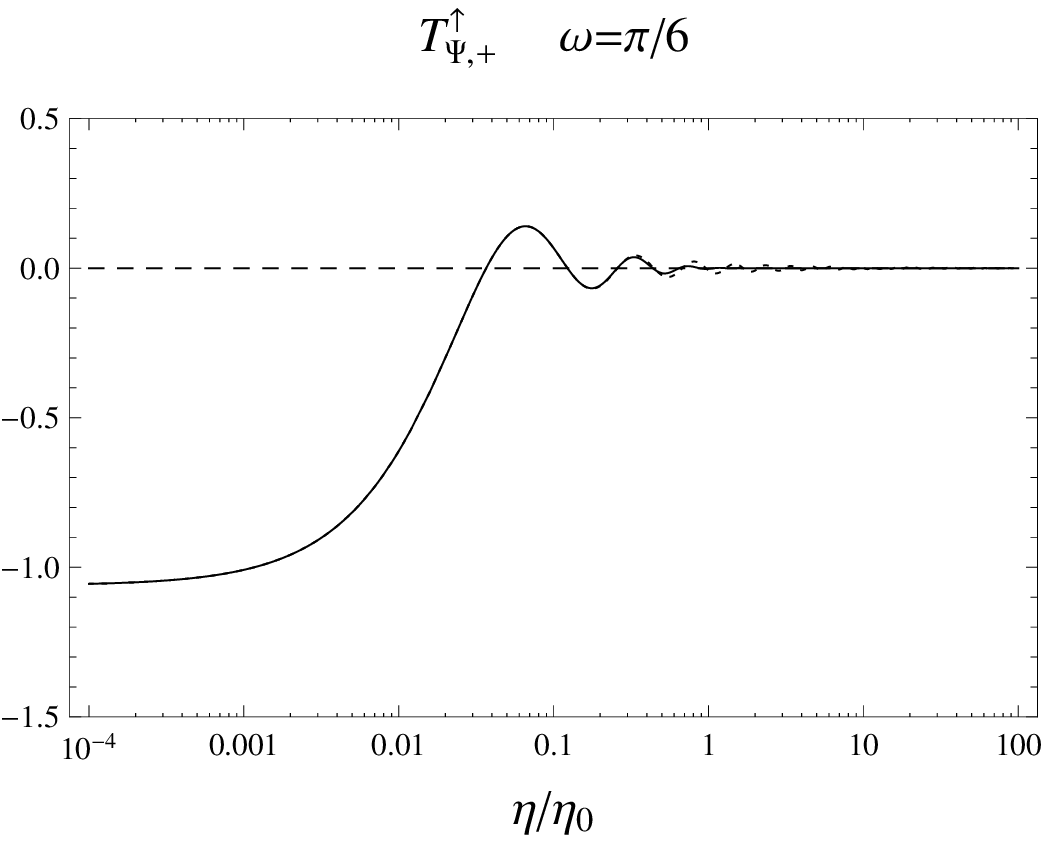}
\caption{Evolution of a mode with $k_1 \eta_0=k_3 \eta_0=10$ and $k_2=0$ in a Kasner universe with
$\varpi=\frac{\pi}{6}$: $T_\times^{\uparrow}$ (left), $T_+^{\uparrow}$ (middle) and $T_\Psi^{\uparrow}$ (right) compared to their asymptotic late time (dashed) and early time
(dotted)  asymptotic forms. The agreement at early times is very good
so that the continuous and dotted curves are hardly distinguishable.} 
\label{fcp1axial}
\end{figure*}

\begin{figure*}
\includegraphics[width=8.5cm]{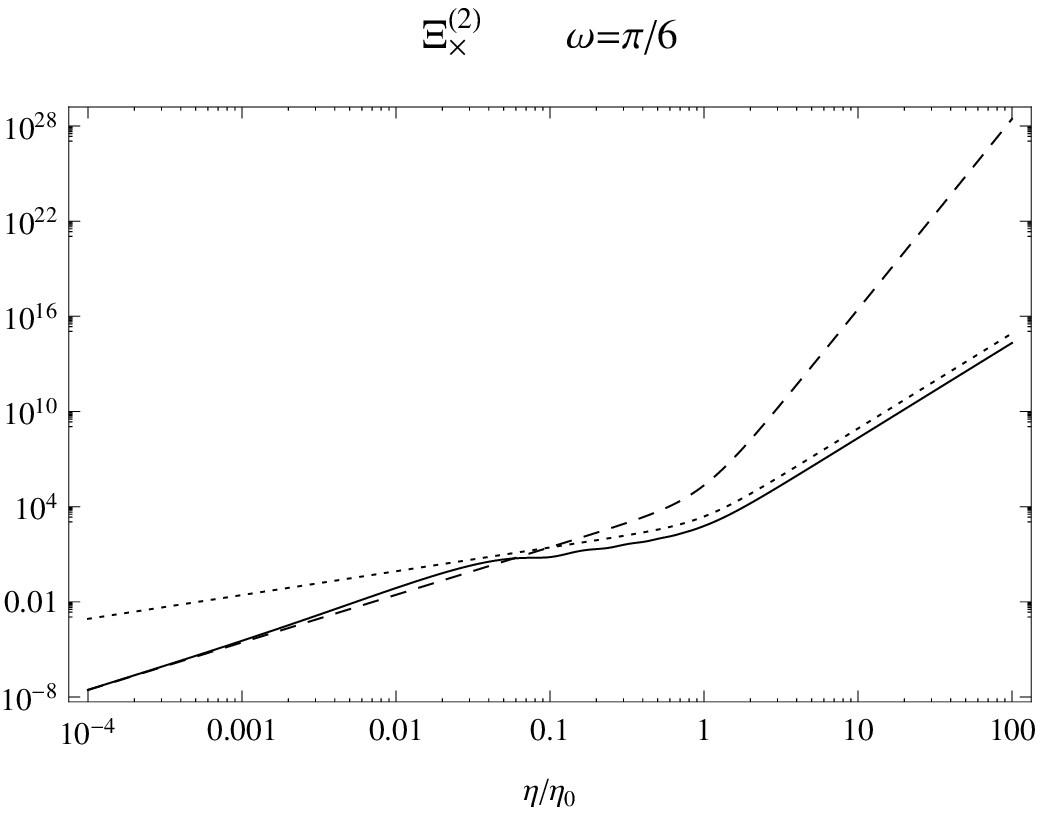}\includegraphics[width=8.5cm]{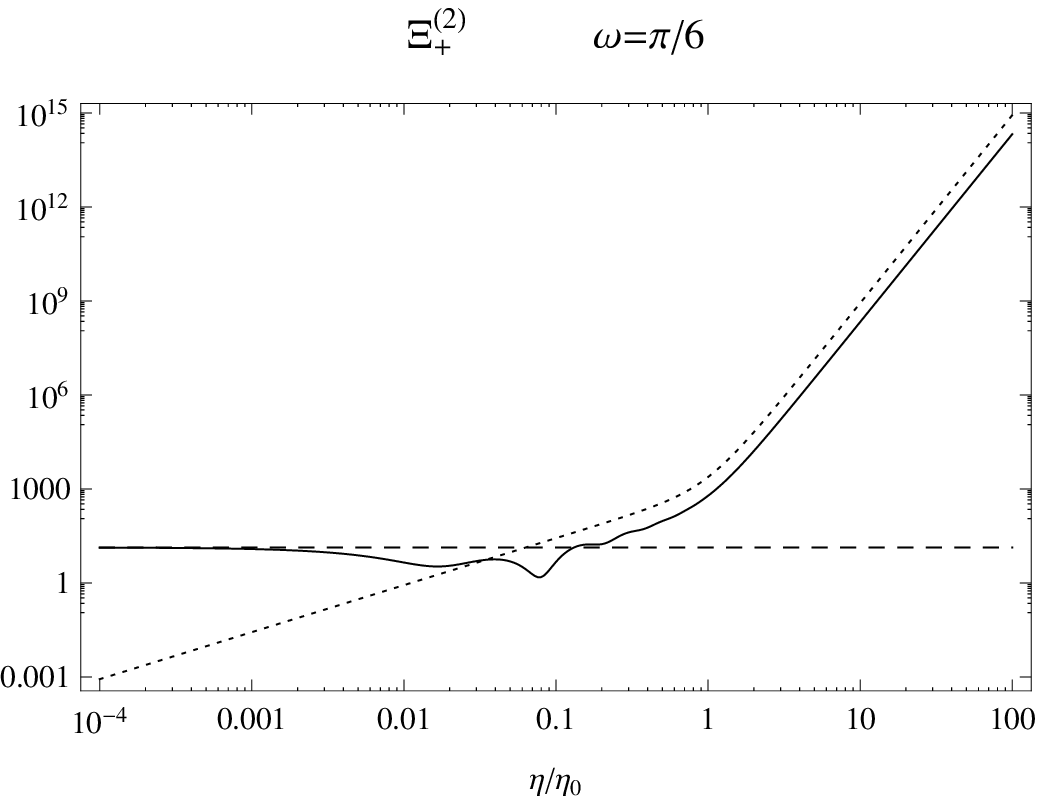}
\caption{Evolution of $\Xi^{(2)}_\lambda({\bf k},\eta)$ for a mode with $k_1 \eta_0=k_3 \eta_0=10$ 
and $k_2=0$ in a Kasner universe with $\varpi=\frac{\pi}{6}$ for the polarisation $\times$ (left) and
$+$ (right) as well as their late time (dashed) and early time (dotted)  asymptotic forms.} 
\label{fcp2axial}
\end{figure*}

Then, we can integrate numerically Eq.~(\ref{mu2axial}) for different modes.
This exact numerical solution is compared in Fig.~\ref{fcp1axial} to the
two asymptotic forms, which demonstrates that the transition is sharp even for
the evolution of the modes.
In Fig.~\ref{fcp2axial}, the full numerical solution of $\Xi^{(2)}_\lambda({\bf k},\eta)$
is compared to the two asymptotic behaviours described in \S\ref{5c} and
\S\ref{5d} . This shows explicitely the validity of the asymptotic
expansions.

\section{Conclusion}

We have performed the stability study  of a generic Kasner spacetime with respect
to linear perturbations using the formalism developed in Refs.~\cite{ppu1,ppu2}.
Since Kasner spacetimes are solutions of the vacuum Einstein equations, one
expects only two degrees of freedom to propagate, which we explicitely demonstrate.
They can be identified with the two polarisations of gravity waves. The extra
scalar and vector components of the metric are then obtained algebraically
from constraint equations. Contrary to a Minkowski or de Sitter spacetime,
they do not strictly vanish because the spatial sections are not isotropic
anymore.

We have shown that for any Fourier mode, the behaviour of the gravity
waves interpolates between two asymptotic regimes, at early and
late times, in which the mode can be considered almost aligned
with a principal axis of the spatial metric. Interestingly, in this
limit the gravity wave equation can be integrated analytically. This allows
to set a lower bound on the square of the Weyl tensor generated by the
perturbations and compare it to the one of the background. We have concluded that at late time the Kasner spacetimes were unstable
with respect to linear perturbations but in the particular case of $\varpi=\frac{\pi}{2}$,
which corresponds to an axisymmetric configuration, product of a two-dimensional
Milne spacetime and a two-dimensional Euclidean spacetime, which actually maps
to one quarter of the Minkowski spacetime. At early time, the conclusion
depends on the modes included in the analysis. If one includes decaying modes,
the Kasner spacetimes are again unstable while stability to linear perturbations
is recovered only if the growing modes are excited. This latter result was
already known since Ref.~\cite{bkl1} and our analysis confirms the
one of Ref.~\cite{emir} that was limited to the particular case
of an axisymmetric configuration with $\varpi=\frac{\pi}{6}$. This
result is also compatible with the amplification of gravity waves
during the shear dominated era of an anisotropic inflationary phase described
by a Bianchi~I spacetime~\cite{ppu2,emir0}. This is amplification of a remnant of the instability
described in this article for a finite Kasner era.

This result has some importance concerning the dynamics of the universe close
to the singularity. On the one hand, the succession of Kasner era in the BKL
formalism and the rotation of the Kasner axes from one era to the other
was studied by assuming homogeneous behaviour, i.e. assuming
the inhomogeneity scale was larger than the mean Hubble patch~\cite{henk}.
The instability at early time (i.e. going toward the singularity) can alter the
conclusions on the approach of the singularity, even though each Kasner era
has a finite duration. In particular, as soon as the effect of the gravity wave is large
enough, the backreaction would have to be taken into account, and it is
not clear that the evolution toward the singularity follows a series
of BKL oscillations. This requires further investigations that go
beyond perturbation theory, see e.g. Ref.~\cite{Deruelle}.
On the other hand, the instability at late time could also be relevant
in models where the pre-inflationary era is described by an anisotropic
era. The gravity waves modes generated during this era can potentially be
observed~\cite{emir2006,ppu2,emir} if the number of e-folds during inflation is not too
large and their presence can modify the onset of inflation.

Besides the speculative phenomenology of the early universe, our result sheds
some light on the peculiarity of the Kasner spacetimes among the
class of homogeneous vacuum solutions of Einstein equations.

\section*{Acknowledgments:}
CP is supported by the STFC (UK) grant ST/H002774/1, and thanks the
Royal Astronomical Society for financial support and the University of
Cape Town for hospitality during part of this work was undertaken. JPU thanks the PNCG for financial support for this project and George Ellis for discussions.
It is a pleasure to thank M. Peloso and E. G\"umr\"uk\c{c}\"uo\u{g}lu for the many
comments they gave us when comparing our results. We also thank Francis Bernardeau, Dick Bond and
Andrei Linde for their kind advices. This work is dedicated to the memory of our friend and colleague
Lev Kofman with whom this project was initiated during his last trip to
Paris in June 2009.

\appendix
\section{Different parameterisations of Kasner exponants}

There exist several ways to parameterize, the coefficients $(p_1,p_2,p_3)$
of the Kasner metric. Because of the two constraints~(\ref{Eqconstraintspi}),
it is in fact a 1-parameter family of triplets that is completely
specified by only one number. Three parameterisations have been 
widely used.
\begin{itemize}
\item{\em Single exponant parameterisation.} As we have seen in \S~\ref{sec2-a}, it is sufficient to
give one of the Kasner exponants to specify the two others. For instance, fixing the 
smallest one, $p_1$, gives that $p_2=1-p_1-p_3$ so that $p_3^2+(1-p_1-p_3)^2+p_1^2=1$,
from which we deduce that
\begin{equation}
 p_2=\frac{1}{2}\left[1-p_1 -\sqrt{(1-p_1)(1+3p_1)} \right],
\end{equation}
\begin{equation}
 p_3=\frac{1}{2}\left[1-p_1 +\sqrt{(1-p_1)(1+3p_1)} \right] .
\end{equation}
The two constraints~(\ref{Eqconstraintspi}) allow to deduce that the $p_i$ satisfy the
following useful relations. First, $(\sum_i p_i)^2=\sum_i p_i ^2$ so that
\begin{equation}
p_1p_2 + p_2p_3 + p_3p_1 =0,
\end{equation}
also conveniently rewritten as
\begin{equation}
 \sum_i p_i^{-1}=0.
\end{equation}
From $(\sum_i p_i)^3=1$, one deduces that
\begin{equation}
\sum_i p_i ^3=1+3\prod_i p_i.
\end{equation}
This tells us that
\begin{equation}\label{Magicrelation123}
p_1(p_1-1)=p_2 p_3
\end{equation}
which implies in particular $p_1^2+p_2^2+ p_1 p_2-(p_1+p_2)=0$, with
similar relations obtained after permutation of the indices $1 \to 2 \to 3 \to 1$.
In terms of the coefficient $q_i$, it leads to the relations
\begin{equation}
2 q_{2,3}= -q_1\pm \Delta_1\,,\qquad \Delta_1\equiv \sqrt{\frac{4}{3}-3 q_1^2}\,.
\end{equation}
In terms of the $Q_i=3/2 q_i$ these relations take the simple form
\begin{equation}
2 Q_{2,3}= -Q_1\pm \sqrt{3(1-Q_1^2)}\,.
\end{equation}
This shows that all the properties of the Kasner spacetime, but also
the evolution of its perturbations when a mode is aligned with an
eigendirection of the shear, are characterized by the choice of a
single Kasner exponent corresponding to that particular direction.
 
\item{\em Angular parameterisation.} As we have seen in \S~\ref{sec2-a}, 
the constraints~(\ref{Eqconstraintspi}) imply that the triplet can be specified
by a choice of an angle $\varpi\in\left[\frac{\pi}{6},\frac{\pi}{2}\right]$ and its expression is given by
Eq.~(\ref{paravarpi}) and we have seen that
\begin{equation}
 \Delta_I\equiv \frac{2}{\sqrt{3}}\vert\cos\varpi_I\vert.
\end{equation}
\item{\em BKL parameterisation.} Refs.~\cite{bkl1,bkl2,bkl3} use the parameterisation
 \begin{equation}
 \hspace{0.7cm} p_1=-\frac{u}{\sigma(u)},\quad p_2=\frac{1+u}{\sigma(u)},\quad p_3=\frac{u(1+u)}{\sigma(u)},
\end{equation}
with $\sigma(u)=1+u+u^2$. It can be obtained~\cite{kasnergen} by solving the
two constraints~(\ref{metricB1}) once having set
\begin{equation}
u = \frac{p_3}{p_2}
\end{equation}
and the condition $p_1\leq p_2\leq p_3$ imposes that
$u\in[1,+\infty[$, with $u=1$ corresponding to 
$\varpi=\frac{\pi}{6}$ and $u\rightarrow+\infty$ to $\varpi=\frac{\pi}{2}$.
\end{itemize}

\section{Axially symmetric case with $\varpi=\frac{\pi}{2}$}\label{secpi2}

The particular case $\varpi=\frac{\pi}{2}$ requires special attention. 
It corresponds to a spacetime with metric
\begin{equation}\label{EqTaubMetric}
 \dd s^2=-dt^2+t^2\dd z^2 + \dd x^2 + \dd y^2,
\end{equation}
which appears to be the product of a two-dimensional Milne
space with a two-dimensional Euclidean space. It is
the Taub representation~\cite{taub} of the Minkowski metric.
With the change of coordinates defined by
\begin{equation}
 (t,x,y,z)\rightarrow (T= t\cosh z,X=x,Y=y,Z=t\sinh z),
\end{equation}
this metric is rewritten as
\begin{equation}
 \dd s^2=-dT^2+ \dd Z^2 + \dd X^2 + \dd Y^2,
\end{equation}
i.e. has a Minkowski metric but only covers the
patch $T^2-Z^2>0$, the upper cone ($T> 0$) corresponding to
an expanding universe while the lower one ($T<0$) 
describes a contracting spacetime.
This explains why $C^2=0$ [Eq.~(\ref{valC})].

The choice of the time slicing plays an important role when
studying the perturbations. In Minkowskian coordinates the spacetime
is explicitely homogeneous and isotropic so that the tensor
modes decouple from the other types of perturbations and 
evolve according to
\begin{equation}
 \left(\partial_T^2 - \Delta_{\lbrace X,Y,Z\rbrace}\right) h_{ij}=0.
\end{equation}
Their amplitude is thus constant and the perturbation of the
square of the Weyl tensor $\delta C^2$ remains constant. Under its Kasner
form, this no more obvious and one needs to check the
consistency with the result in Minkowskian coordinates
\begin{equation}\label{eb5}
\delta^{(2)} C^2_\lambda = 8 \left[k^2 \left({\frac{\partial 
T^{\uparrow}_\lambda}{\partial \eta}}\right)^2+ k^4 {T^{\uparrow}_\lambda}^2\right]
\end{equation}
where the two modes have constant amplitude (so that the distinction
between growing and decaying is no more relevant).

\subsection{Description of the perturbation from a
Kasner point of view}

\subsubsection{Gravity waves equation}

We follow the general formalism of this article to study the perturbation
in the particular case
\begin{equation}
 q_1=q_2=-\frac{1}{3},\qquad
 q_3=\frac{2}{3}
\end{equation}
so that
\begin{equation}
 \Delta_1=\Delta_2=1,\qquad
 \Delta_3=0.
\end{equation}
It implies that
\begin{equation}
 k^2=k_3^2\left(\frac{\eta}{\eta_0}\right)^{-2} + k_\perp^2\left(\frac{\eta}{\eta_0}\right).
\end{equation}
Obviously, this case can be deduced from the case $\varpi=\frac{\pi}{6}$
by $q_i\rightarrow-q_i$ so that the Euler angles are again
\begin{equation}
 \tan\gamma=\frac{k_1}{k_2},
 \qquad
  \alpha=0.
\end{equation}
It follows that the expressions for the
shear components are also given by Eqs.~(\ref{s1axial}-\ref{s3axial})
but with all signs changed, ie. $\Spar\rightarrow-\Spar$ etc.
The evolution of the gravity waves is dictated by Eq.~(\ref{mu2axial}) with
\begin{equation}
 \Spar =-\sin^2\beta+2 \cos^2\beta,
 \end{equation}
\begin{equation}
 \Svun =-3 \sin\beta \cos\beta,\qquad
 \Svdeux=0
\end{equation}
and
\begin{eqnarray}
 \Stplus = \frac{3}{\sqrt{2}}\sin^2\beta,\qquad
 \Stcross = 0.
\end{eqnarray}
The only difference concerns the Euler angle $\beta$ which is now given by
\begin{equation}
 \cos\beta = \frac{k_3}{k}\left(\frac{\eta}{\eta_0}\right)^{-1},
 \quad
 \sin\beta = \frac{k_\perp}{k}\left(\frac{\eta}{\eta_0}\right)^{\frac12}.
\end{equation}

\subsubsection{Late time behaviour of the perturbations}

In the limit $\eta\rightarrow+\infty$, $k\sim k_\perp(\eta/\eta_0)^{\frac12}$
so that $x\sim k_\perp\eta_0(\eta/\eta_0)^{\frac32}$. This
implies that
\begin{equation}
 \cos\beta \sim \frac{k_3}{k_\perp}\left(\frac{\eta}{\eta_0}\right)^{-\frac32},
 \quad
 \sin\beta \sim1
\end{equation}
$\Spar\sim -1$, $\Svun\sim - 3\cos\beta$ and $\Stplus\sim3/\sqrt{2}$ so that
we conclude
\begin{equation}
X \sim 
\frac{E_+}{\sqrt{2}},\quad
 \Phi_a \sim\frac{3{\rm i}\sqrt{2}}{k_\perp\eta_0}\frac{k_3}{k_\perp}  \left(\frac{\eta}{\eta_0}\right)^{-3}\left(\begin{array}{c}
 \frac{1}{2}E_+ \\
  E_\times
 \end{array}\right).
\end{equation}
The solution of the gravity wave equation is then explicitely given by
\begin{eqnarray}
 E_\lambda \sim
 \sqrt{\frac{3}{\pi x}}A_{\lambda}^{(\infty)}
 \cos\left[\frac{2}{3}x+\varphi^{(\infty)}_\lambda \right],
\end{eqnarray}
for the two polarisations, and where the growing and decaying modes
are combined together, leading to a redefinition of the phase.

Since $\bar C^2=0$, we cannot use the variable $\Xi$ anymore and thus work 
directly with $\delta^{(2)} C^2$. In the limit $\eta \to \infty$, we obtain 
\begin{eqnarray}
\delta^{(2)} C^2_\lambda({\bf k},\eta) &\to& \frac{8 x^2}{S^4 \eta^4}
\left[\left({\frac{\partial T^{\uparrow}_\lambda}{\partial \ln
        \eta}}\right)^2+ x^2
  {T^{\uparrow}_\lambda}^2\right]\nonumber\\
&\to&\frac{24 x^3 }{\pi \eta^4 S^4}\propto\eta^{-\frac{3}{2}}.
\end{eqnarray}
It follows that $\delta^{(2)} C^2_\lambda$ decays at late time so that the spacetime is stable
against the perturbations. Note that this is the same behaviour as for the case $\varpi\not=\frac{\pi}{2}$, but the main difference arises from the fact that $\bar C^2=0$ since in the general case
$\delta^{(2)} C^2$ decreases with time but slower than $\bar C^2$ so that the effect of the perturbations
dominate over the background quantity at late time.
The stability of the Taub metric~(\ref{EqTaubMetric}) was studied in
Ref.~\cite{taubstab} with the same conclusions as our analysis.

\section{Bessel functions}

The differential equation
\begin{equation}\label{trucjp1}
 u'' + \left[\left(\beta\gamma x^{\gamma-1}\right)^2+\left(\frac14-\nu^2\gamma^2\right)\frac{1}{x^2} \right] u =0
\end{equation}
has the general solution
\begin{equation}\label{trucjp2}
 u(x) = \sqrt{x}Z_{\nu}\left[\beta x^\gamma \right],
\end{equation}
where $Z_\nu$ is a linear combination of a Bessel function of the first kind ($J_\nu$) and of the
second kind ($N_\nu$ or Newmann function).

We have that in $x\sim\infty$,
\begin{equation}\label{besselJinf}
 J_\nu(x) \sim \sqrt{\frac{2}{\pi x}}\cos\left[x -\nu\frac{\pi}{2}- \frac{\pi}{4} \right]
\end{equation}
while in $x\sim0$,
\begin{equation}\label{besselJ0}
 J_\nu(x) \sim \frac{1}{\Gamma(1+\nu)}\left(\frac{x}{2}\right)^\nu.
\end{equation}

We have that in $x\sim\infty$,
\begin{equation}\label{besselNinf}
 N_\nu(x) \sim \sqrt{\frac{2}{\pi x}}\sin\left[x -\nu\frac{\pi}{2}- \frac{\pi}{4} \right]
\end{equation}
while in $x\sim0$,
\begin{eqnarray}
 N_0(x)& \sim& \frac{2}{\pi}\left[\ln\left(\frac{x}{2}\right)+\gamma_E\right]\label{besselN0a}\\ 
 N_\nu(x) &\sim & - \frac{\Gamma(\nu)}{\pi}\left(\frac{2}{x}\right)^\nu.\label{besselN0b}
\end{eqnarray}


\end{document}